\documentclass[aps,prd,superscriptaddress,amsmath,amssymb,twocolumn,preprint,nofootinbib,10pt]{revtex4-2}
\usepackage{graphicx} 
\usepackage{dcolumn}
\usepackage{bm}
\usepackage[bookmarks=false]{hyperref}
\usepackage{color}
\usepackage{multirow}
\usepackage{balance}
\usepackage{orcidlink}
\usepackage{enumerate}

\definecolor{mygreen}{RGB}{0,155,0}

\def\cb#1{\textcolor{black}{#1}}
\def\c#1{\textcolor{black}{#1}}


\begin{document}

\title{Effects of dynamical capture on two equal-mass nonspinning black holes}
\author{Jorge L. Rodríguez-Monteverde\,\orcidlink{0009-0000-2965-0726}}\email[]{jorge.lopezr@estudiante.uam.es}
\affiliation{Universidad Autónoma de Madrid, Cantoblanco 28049 Madrid, Spain}

\author{Santiago Jaraba\,\orcidlink{0000-0002-4759-143X}}\email[]{santiago.jaraba-gomez@astro.unistra.fr}
\affiliation{Observatoire Astronomique de Strasbourg, CNRS, Université de Strasbourg, 11 rue de l'Université, 67000 Strasbourg, France}
\author{Juan García-Bellido\,\orcidlink{0000-0002-9370-8360}}\email[]{juan.garciabellido@uam.es}
\affiliation{Instituto de Física Teórica UAM-CSIC, Universidad Autónoma de Madrid, Cantoblanco 28049, Madrid, Spain}

\date{\today}
\preprint{IFT-UAM/CSIC-25-116}

\begin{abstract}
Dynamical captures of black holes are unique events that provide an exceptional opportunity to probe the strong-field regime of gravitational physics. In this article, we perform numerical relativity simulations to study the events of dynamical capture of two equal-mass nonspinning black holes. We consider a suite of scenarios within a range of initial linear momenta ($p/M=0.095-0.75$) and incidence angles ($\theta=6.36^\circ-2.83^\circ$), and study the emitted Weyl scalar ($\Psi_4$) of each case, as well as the spins and masses of the black holes before and after they merge. We provide a simple analytical model which accurately fits the gravitational-wave emission. We study the dependence of the time interval between the capture and the merger emissions with respect to the incidence angle, which can be well parametrized by a first-order divergent behavior, allowing us to find the angle that separates a scattering event from a dynamical capture. We also find that, in general, the parameters that model the first emission can be well described by linear or exponentially decaying functions in terms of the incidence angle, while others display more complex behaviors that offer valuable insights into the nature of these events.
\end{abstract}
\maketitle
\section{Introduction}
Dynamical capture (DC) and close hyperbolic encounters (CHEs) of black holes (BHs) have recently gained increasing attention as key processes in dense astrophysical environments. CHEs, in particular, are of growing interest due to their potential implications for the dynamics and evolution of primordial black holes (PBHs) in dense clusters~\cite{Garcia-Bellido:2017qal, Garcia-Bellido:2017knh, Nelson:2019czq, Jaraba:2021ces, Garcia-Bellido:2021jlq, Morras:2021atg, Rodr_guez_Monteverde_2025, Fontbute:2024amb}. On the other hand, a DC can be understood as the limiting case in which a hyperbolic encounter radiates sufficient energy and angular momentum through gravitational waves (GWs) to bind two initially unbound BHs into a merging system. Together with pure scattering events, these scenarios are expected to play a central role in the dynamics of BH clusters, influencing their long-term evolution and the properties of the gravitational wave signals they generate~\cite{Nagar_2021, Albanesi_2025, Trenado:2025ccf}.

Accurately modeling these interactions is crucial for the development of precise N-body simulations of dense BH clusters. Such simulations will allow us to predict the resulting mass and spin distributions of black holes and assess the detectability of the GWs they produce~\cite{Siles:2024yym}. However, the observational prospects for identifying DC events with current ground-based detectors such as those of the LIGO-Virgo-KAGRA (LVK) Collaboration~\cite{LIGOScientific:2014pky,VIRGO:2014yos,KAGRA:2020tym} remain limited. Detectability is hampered by the requirement of long-duration measurements and high-precision sensitivity. Nonetheless, as detector sensitivity improves, it is plausible that DC signatures may be identified in LVK~\cite{OLeary:2016ayz,Albanesi_2025, Liu:2020ufc}. In fact, the event GW190521 was proposed as being the first detection of a GW signal coming from a dynamical capture of nonspinning black holes, preferred over a quasicircular spin-precessing merger~\cite{Gamba_2022}.

In this work, we use numerical relativity (NR) to run simulations of initially nonspinning, equal-mass black holes ($q\equiv m_2/m_1=1$). We leave the inclusion of spin and mass-ratio asymmetries to future studies. This choice is motivated by the expectation that PBHs form with low initial spins~\cite{Chiba_2017}, with their angular momenta subsequently altered through repeated dynamical interactions, most prominently via CHEs, as shown in previous studies~\cite{Nelson:2019czq,Jaraba:2021ces,Rodr_guez_Monteverde_2025}.

This paper is organized as follows: in Sec.~\ref{sec: Grid, Weyl, BH}, we present some basic theoretical aspects that will be relevant for our discussion, and also describe the numerical setup we use and our choice of initial conditions; in Sec.~\ref{Sec: Num. res. Weyl}, we discuss the general behavior of our simulations, the concrete details of the used waveform model, and the results this phenomenological model provides; in Sec.~\ref{sec: Num. res. BH}, we analyze many relevant properties related to the mass and spin of the initial and final black holes; and lastly, in Sec.~\ref{sec: Conclusions}, we summarize our findings and outline future directions.

Finally, the results in this work will be presented using geometrized units, $G=c=1$, the usual convention for NR simulations.

\section{Preliminary framework}
\label{sec: Grid, Weyl, BH}

To properly discuss the phenomenology that arises from dynamical captures, we must first introduce some fundamental aspects that are necessary to understand and build a DC of BHs simulation. Firstly, we briefly present some theoretical aspects such as the outgoing gravitational radiation, the black hole apparent horizons and the measurement of BH spins; and secondly, we introduce the numerical framework: the grid setup, the general initial BH configuration and the selection of initial conditions used in our simulations.

\subsection{Weyl scalar}

Gravitational-wave radiation can be extracted from the \emph{Newman-Penrose formalism} via the Weyl scalar $\Psi_4$, which encodes the outgoing gravitational radiation in terms of a null tetrad basis~\cite{Newman:1961qr, Maggiore:2018sht}, as follows:
\begin{equation}
    \Psi_4 = C_{\mu\nu\rho\sigma} \, n^\mu \overline{m}^\nu n^\rho \overline{m}^\sigma,
\end{equation}
where $C_{\mu\nu\rho\sigma}$ is the Weyl tensor, $n^\mu$ is one of the null basis vectors, and $\overline{m}^\mu$ is the complex conjugate of the corresponding tetrad vector $m^\mu$.

The quantity $\Psi_4$ captures the curvature perturbations associated with outgoing gravitational radiation and is directly related to the observable gravitational-wave strain. The complex combination of the two polarization states, $h_+$ and $h_\times$, can be obtained from $\Psi_4$ through a double time integration,
\begin{equation}
    h_+(\vec{x},t) - i h_{\times}(\vec{x},t)
    = \int_{-\infty}^t \! du \int_{-\infty}^{u} \! dv \, \Psi_4(\vec{x},v),
    \label{Eq: Weyl-strain2}
\end{equation}
where both the strain components and the Weyl scalar can be expanded in terms of their spherical harmonic modes in the following manner:
\begin{equation}
    h_{+,\times}(\vec{x},t) = \sum_{l=2}^\infty \sum_{m=-l}^{+l} h^{lm}_{+,\times}(r,t) Y^{lm}_{-2}(\theta, \phi).
    \label{Theory:strain}
\end{equation}
In practice, extracting the quantity $\Psi_4$ from NR simulations provides a robust and gauge-invariant measure of gravitational radiation. For both clarity and numerical accuracy, our analysis focuses on the dominant quadrupolar mode, $\Psi_4^{(2,2)}$, which typically carries the majority of the emitted energy in binary black hole (BBH) coalescences.

\subsection{Apparent horizons}
\label{Subsec. app-hor}

The identification of BH horizons plays a central role in accurately describing the dynamics of the full BBH system within an NR simulation. In dynamical spacetimes, however, the event horizon (the null hypersurface that defines the boundary of the BH region of a spacetime) is a global property. This makes it impractical for real-time numerical tracking, since its location depends on the entire future evolution of the spacetime.

Instead, apparent horizons provide a local and physically meaningful notion of a black hole boundary during dynamical evolution. An apparent horizon is defined as a closed two-surface where the expansion of outgoing null geodesics vanishes, i.e., a surface that locally separates outgoing and ingoing light rays. Mathematically, it satisfies
\begin{equation}
    q^{ij}(K_{ij} - D_i s_j) = 0,
    \label{eq: app. hor}
\end{equation}
where $s_i$ is the spacelike unit normal to the horizon surface, $D_i$ denotes the covariant derivative compatible with the spatial metric $\gamma_{ij}$, $q_{ij} = \gamma_{ij} - s_i s_j$ is the induced two-metric on the surface, and $K_{ij}$ is the extrinsic curvature~\cite{Altas_2022}.

Once the apparent horizon is located, its area $A$ can be computed, which in turn allows us to define the horizon mass~\cite{Brandt_1996}, as follows:
\begin{equation}
    m_H = \sqrt{\frac{A}{16\pi} + \frac{4\pi \c{S}^2}{A}},
    \label{eq: hor mass}
\end{equation}
where $\c{S}$ is the angular momentum of the black hole.

To interpret this relation, it is convenient to introduce two key mass definitions. The Arnowitt–Deser–Misner (ADM) (or horizon) mass, $m \equiv m_H$, represents the total energy content of the isolated system in an asymptotically flat spacetime. The irreducible mass,
\begin{equation}
    m_{\text{irr}} \equiv \sqrt{\frac{A}{16\pi}},
    \label{eq: irred-area}
\end{equation}
corresponds to the portion of the black hole’s mass that cannot be reduced by any classical process, even those capable of extracting rotational energy (such as the Penrose process or black hole mergers)~\cite{ruffini2024roleirreduciblemassrepetitive}.  

In terms of these quantities, Eq.~\eqref{eq: hor mass} can be rewritten as
\begin{equation}
    m_H = \sqrt{m_{\text{irr}}^2 + \frac{\c{S}^2}{4m_{\text{irr}}^2}},
    \label{eq: ADM-irreducible}
\end{equation}
which elegantly expresses the total black hole mass as the sum of its irreducible (nonextractable) and rotational energy contributions.

\subsection{Measuring spin}
\label{subsec: spin measurements}
The spin of a black hole in a dynamical spacetime is a gauge-dependent quantity, and therefore admits several distinct definitions. Only in the limit where BHs are well separated and approximately stationary do all spin measures coincide. One of the most practical ways to estimate BH spins in NR simulations is through the geometric properties of their apparent horizons.  

The spin can be inferred from the shape of the apparent horizon via the relation~\cite{Alcubierre_2005}
\begin{equation}
    \frac{C_p}{C_e} = \frac{1+\sqrt{1-\chi^2}}{\pi}
    \,E\!\left(\frac{-\chi^2}{(1+\sqrt{1-\chi^2})^2}\right),
    \label{eq: manual, spins}
\end{equation}
where $C_p$ and $C_e$ denote the polar and equatorial circumferences (measured along their respective geodesics), $\chi=a/m = \c{S}/m^2$ is the dimensionless spin parameter (more thoroughly defined in previous references such as Refs.~\cite{Rodr_guez_Monteverde_2025,Jaraba:2021ces, Gerosa_2023}), and $E(x)$ is the complete elliptic integral of the second kind, defined as
\begin{equation}
    E(x) = \int_0^{\pi/2} d\theta \, \sqrt{1 - x \sin^2\theta}.
\end{equation}

In our simulations, this computation is performed automatically by the \texttt{QuasiLocalMeasures} thorn~\cite{Dreyer:2002mx}, which provides a consistent framework for evaluating quasilocal quantities such as mass, spin, and angular momentum directly from the horizon geometry.

\subsection{Grid properties and initial black hole configuration}
\label{Subsec: setup}

The simulations were performed using the Einstein Toolkit~\cite{Loffler:2011ay,EinsteinToolkit:web}. In particular, we employed the Cactus Computational Toolkit~\cite{Goodale:2002a,Cactuscode:web} as the core infrastructure, with adaptive mesh refinement handled by \texttt{Carpet}~\cite{Schnetter:2003rb,CarpetCode:web}. The initial puncture data were generated using \texttt{TwoPunctures}~\cite{Ansorg:2004ds,Paschalidis:2013oya}, and the spacetime evolution was carried out using the Baumgarte–Shapiro–Shibata–Nakamura formulation implemented in \texttt{McLachlan}~\cite{Brown:2008sb,Kranc:web,McLachlan:web}. To track the apparent horizons, we utilized the \texttt{AHFinderDirect} thorn~\cite{Thornburg:2003sf,Thornburg:1995cp}, while the \texttt{QuasiLocalMeasures} thorn~\cite{Dreyer:2002mx} was used to compute black hole spins. Lastly, the Weyl scalar $\Psi_4$, used for gravitational-wave extraction, was obtained through the \texttt{WeylScal4} thorn~\cite{Zilhao:2013hia}.

The initial setup of our simulations is identical to that described in the first sections of Ref.~\cite{Jaraba:2021ces}. This means we consider equal-mass binary black holes, $m_1 = m_2 \equiv m$, with symmetric linear momenta, $|\vec{p}_1| = |\vec{p}_2| \equiv p$, as well as no spins, $\vec{S}_1=\vec{S}_2=0$. Working in geometrized units, all quantities such as length, time, and momentum are expressed in units of mass. As is standard in NR, we fix the total mass $M \equiv m_1 + m_2 = 2m = 1$ in the simulations. Therefore, results for a generic total mass $M$ can be obtained by rescaling the dimensionless quantities, which are presented as $t/M$, $p/M$, $x/M$, etc. Working in these units and following the literature, the initial \c{coordinate} distance between the BHs has been set to $d=100M$.

Regarding the computational setup, we follow the same grid configuration as in Refs.~\cite{Jaraba:2021ces,Rodr_guez_Monteverde_2025, Nelson:2019czq}. The grid hierarchy consists of refinement levels with half-lengths of $0.75 \times 2^n$ for $n = 0, 1, \ldots, 6, 8, 9, 10$, and grid spacings of $2^n \times \Delta x_{mr}$ for $n = 0, 1, \ldots, 9$, where $\Delta x_{mr}$ denotes the resolution of the finest grid. In our simulations, we employ a single resolution, $\Delta x_{mr} = (3/200)M$, which in previous works was referred to as \textit{medium} resolution~\cite{Jaraba:2021ces,Rodr_guez_Monteverde_2025}.

Based on these references, and on explicit cross-checks performed here, we find that (with regards to the induced spins) differences between the resolution used here and higher ones remain below $0.6\%$. Other relevant variables have also been checked to have good accuracy with differences below $1\%$. Given that the maximum induced spins in our nonspinning case reach only $\chi \sim 0.14$ (in agreement with previous results~\cite{Rodr_guez_Monteverde_2025}), we conclude that medium resolution is sufficient to obtain accurate and robust results. Nevertheless, in some simulations dealing with highly relativistic initial conditions (more specifically $p/M=0.75$), the formed BH after the merger was observed to have large discrepancies \c{with higher resolutions} in masses and spins, which is the reason we discarded these concrete values for these simulations, as \c{indicated} in Sec.~\ref{sec: Num. res. BH}.

\subsection{Choice of initial conditions}
\label{Sec: ini-cond-DC}

We perform a series of simulations using the same setup as in Ref.~\cite{Jaraba:2021ces}, considering equal-mass binaries. The values of the \c{initial} dimensionless linear momentum $p/M$ \c{are set to six different values between 0.095 and 0.75, while the incidence angles $\theta$ aim to cover, for each of these momenta, the full range of cases corresponding to dynamical capture events.} Table~\ref{Tab: Cases} lists the number of simulations performed for each $p/M$, alongside with the ranges of incidence angles \c{corresponding to dynamical capture events}. We also provide fitted impact parameters, determined as explained in Appendix~\ref{app: impact pars}.

\c{In order to test the limits of the incidence angle range, simulations with lower and higher angles were also performed, ultimately being excluded from Table~\ref{Tab: Cases} as they correspond to either pure mergers (lower angles) or hyperbolic encounters (higher angles). In this work, we consider an event to be a pure merger when its Weyl scalar amplitude shows a single local maximum on its emission peak. On the other hand, an event is considered as a hyperbolic encounter when both black holes separate from each other after the initial burst, showing no signs of getting closer again up to BH separations of $d\gtrsim 70M$. On the lower end of these ranges of incidence angles, the difference from the last DC simulation to the first merger ranges from $\delta\theta=0.009^\circ$ for the lowest momentum to $0.003^\circ$ for the highest one. On the upper end, the distance from the last DC to the first hyperbolic encounter ranges from $0.023^\circ$ to $0.006^\circ$.}

\begin{table}[htbp]  
\caption{Considered values of $p/M$ along with their respective ranges of $\theta$ and the fitted ranges of impact parameters, $b/M$. The value $N_S$ represents the number of simulations that have been computed for the given scenario.}
\centering 
\renewcommand{\arraystretch}{1.5}
\setlength{\tabcolsep}{3.3pt}
\begin{tabular}{|c|c|c|c|c|}
\hline  
 $p/M$ & $\theta$ (deg) &$b/M$ & $N_S$ \\ [0.5ex]  
\hline   
$0.095$  &$ 6.\c{188}-6.3\c{57}$ & $10.7\c{2}-11.02$ & \c{10} \\ [0.5ex] 
\hline   
$0.1225$  &$ 5.04\c{8}-5.1\c{39}$ & $8.76-8.91$ & 10 \\ [0.5ex] 
\hline   
$0.245$  &$ 3.29\c{4}-3.32\c{4}$ & $5.71-5.77$ & 12  \\ [0.5ex]
\hline
$0.3675$  &$2.90\c{2}-2.9\c{28}$ & $5.03-5.07$ & 10 \\ [0.5ex]
\hline
$0.49$  &$2.83\c{3}-2.8\c{56}$ & $4.90-4.94$ & 9\\ [0.5ex]
\hline
$0.75$  &$3.05\c{4}-3.09\c{7}$ & $5.26-5.33$ & 9\\
\hline
\end{tabular}  
\label{Tab: Cases}
\end{table} 

From Table~\ref{Tab: Cases}, we observe that the ranges of incidence angles and impact parameters that lead to capture become wider for smaller BH momenta. This is consistent with the expectation that slower-moving BHs require less precise alignment (i.e., larger $\theta$) to undergo capture. Conversely, higher-momentum BHs require smaller impact parameters and incidence angles for capture. Interestingly, for the scenario with $p/M=0.75$,\footnote{Analogously to Ref.~\cite{Rodr_guez_Monteverde_2025}, we use the term ``case'' to refer to a specific simulation with fixed initial conditions $\{p/M,\theta\}$, and the term ``scenario'' for the ensemble of cases corresponding to a given $p/M$.} the range of incidence angles increases again; this peculiar behavior will be discussed in the next section. 

In the following analysis, we study the evolution of BH properties throughout the interaction and after the merger, and characterize the corresponding GW signals. For quantities such as mass and spin, we focus on the postmerger regime, when these parameters reach stable asymptotic values (as indicated in previous works~\cite{Jaraba:2021ces,Rodr_guez_Monteverde_2025}, and as will be seen in the plateau parts of figures in Sec.~\ref{subsec: gen. beh-DCs}). For the GW analysis, we focus on the regions of the signal where the Weyl scalar exhibits pronounced activity, examining the amplitude, frequency, and phase shift of each emission.

\section{Numerical results}
\label{Sec: Num. res. Weyl}
\subsection{General behavior of DCs}
\label{subsec: gen. beh-DCs}

As discussed in the Introduction, DCs behave similarly to CHEs, with the key difference being that the GW emission produced during the CHE is strong enough to bind the system, ultimately leading to a merger. 

\begin{figure}[htbp]
    \centering
    \includegraphics[width=\columnwidth]{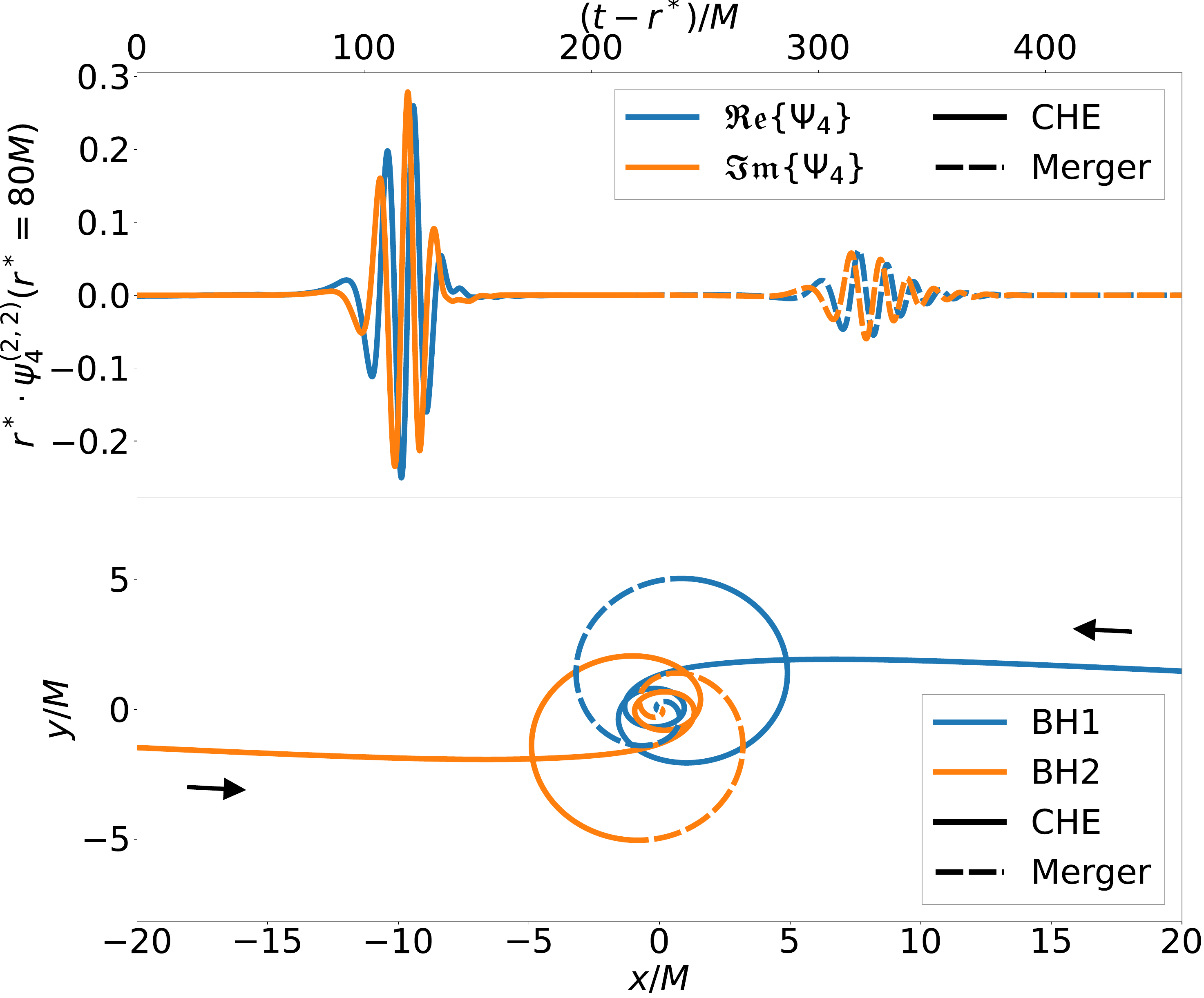}
    \caption{Upper panel: evolution of the $(2,2)$ multipole of the Weyl scalar measured at a detector located at a distance $r^*=80M$ from the c.m. Lower panel: trajectories of both BHs, initially hyperbolic. The arrows qualitatively indicate the initial momentum $\vec{p}$ of each BH. The solid and dashed lines roughly show when the BHs are in their CHE or merger \c{stages}, respectively. The initial conditions are $p/M=0.49$, $\theta=2.847^\circ$.}
    \label{psi+tray}
\end{figure}

In Fig.~\ref{psi+tray} (upper panel), we show the Weyl scalar for a specific choice of $p/M$ and $\theta$, together with the trajectories of both BHs shown in the lower panel. Two distinct GW emissions are clearly visible: the first corresponds to the CHE, which binds the orbit, and the second to the final merger of the BHs. As seen in the lower panel, the initial close approach during the CHE produces a wide, temporary orbit lasting approximately $ 200M$, followed by a nearly head-on merger.

A relevant aspect to investigate in the following is how the initial conditions influence the amplitude and structure of these emissions. We expect that larger initial momenta lead to stronger GW emission during the CHE, since higher velocities imply more energetic and abrupt encounters with stronger frame-dragging effects. This trend is also tied to the impact parameter: higher values of $p/M$ require smaller incidence angles to result in capture, thereby producing more intense GW bursts. This behavior will become clearer in the next section.

\begin{figure}[htbp]
    \centering
    \includegraphics[width=\columnwidth]{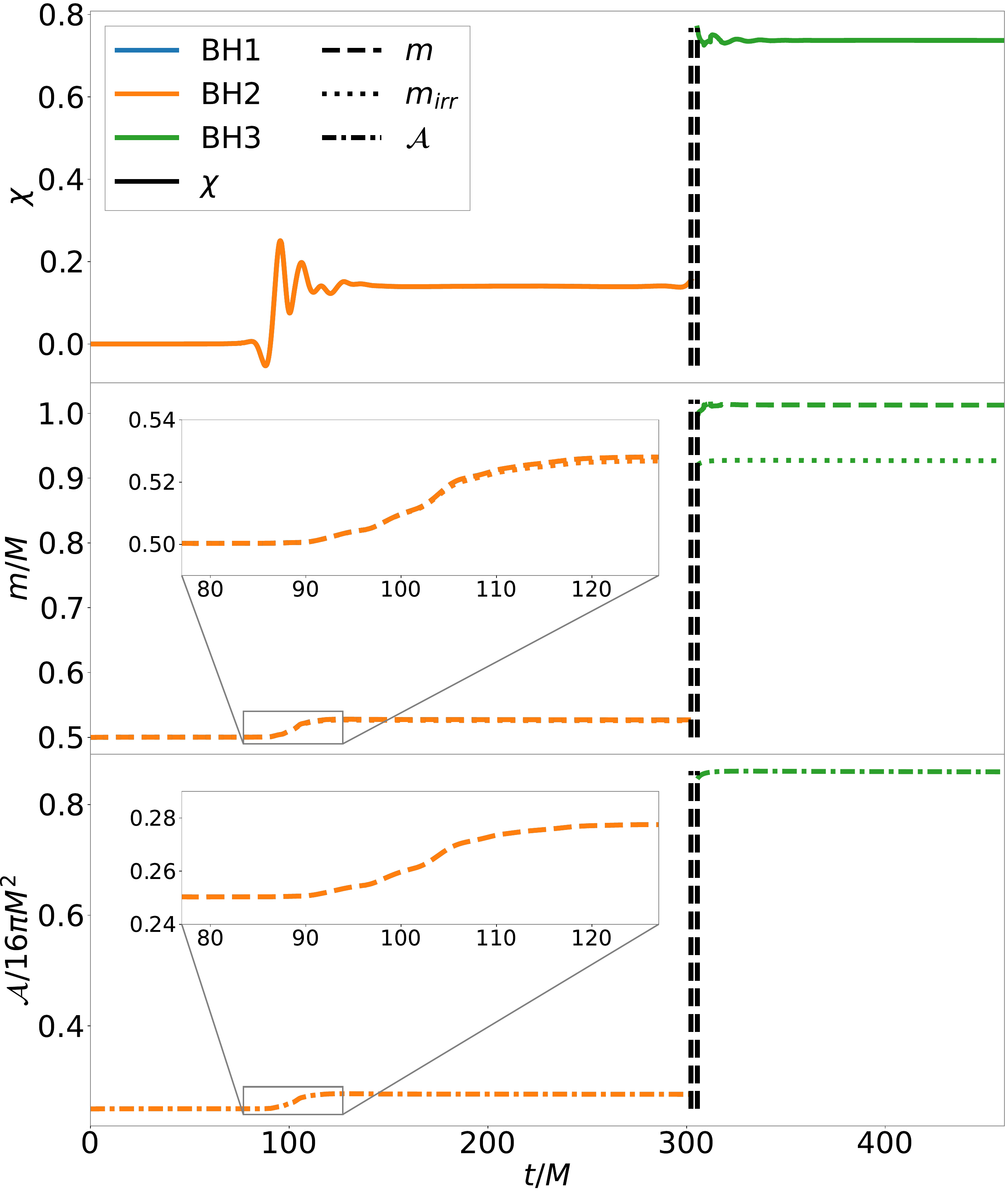}
    \caption{Upper panel: evolution of the dimensionless spin parameter $\chi$ for each BH and for the final remnant. Middle panel: evolution of the ADM and irreducible masses of both BHs and of the merged remnant. Lower panel: evolution of the area ${\cal A}$ for both BHs and the final one. Insets in the two lower panels are shown around the CHE \c{stage}. The initial conditions are $p/M=0.49$, $\theta=2.847^\circ$. The black vertical dashed lines indicate the moment of merger. Both BHs evolve in the same way, as their initial conditions are identical (see Sec.~\ref{Subsec: setup}).}
    \label{chi+mass}
\end{figure}

In Fig.~\ref{chi+mass}, we present the time evolution of the spin $\chi$, ADM mass, and irreducible mass for each BH, as well as for the final remnant. Different colors are used to distinguish the final BH from the initial ones. Because we work in the c.m. frame with equal-mass BHs, both initial BHs evolve identically. During the CHE \c{stage}, there is a noticeable spin-up effect, consistent with previous findings for close hyperbolic interactions~\cite{Jaraba:2021ces, Rodr_guez_Monteverde_2025}.

In this example, the spin increases by approximately $ 0.14$ during the CHE, while the ADM mass grows by about $5\%$. Both of these effects (tidal torquing and heating, respectively) are due to horizon absorption of backreacted GWs during the close encounter~\cite{Chiaramello:2024unv, PhysRevD.108.044039}. Interestingly, the mass of the final BH slightly exceeds the combined pre-CHE masses of the two initial BHs, reaching about $1.01M$. This shows that DCs can yield remnants with total mass greater than the initial sum, even after accounting for energy radiated away as GWs during the merger.

Furthermore, the lower panel of Fig.~\ref{chi+mass} reveals that the horizon area of each BH also increases during the CHE. This is expected, since from Eq.~\eqref{eq: irred-area} the area is directly related to the irreducible mass. The resulting remnant BH possesses an area larger than the sum of the progenitors, in accordance with Hawking’s area law~\cite{PhysRevLett.26.1344, KAGRA:2025oiz, PhysRevLett.25.1596}. This behavior will be systematically verified for all simulations in Sec.~\ref{sec: Num. res. BH}.

As previously discussed in this work and in other studies~\cite{Nelson:2019czq,Jaraba:2021ces}, DCs and pure CHEs exhibit qualitatively similar dynamics, with DCs ultimately leading to mergers that produce a second, prominent GW burst. This can also be visualized from the signal Q-plots, as shown in Appendix~\ref{app: Q-plots}. In the following section, we determine the critical incidence angle that marks the transition between the pure CHE regime and the DC regime.

\subsection{Weyl scalar phenomenological model}
\label{Subsec: Study of the Weyl}

Our next goal is to model the time evolution of \c{the $l=m=2$ mode of} the Weyl scalar $\Psi_4$ for each simulation. To do so, we adopt a simple two-stage fitting procedure:

\begin{enumerate}[i.]
    \item First emission (CHE \c{stage}): modeled as a \textit{\c{Gaussian-modulated complex exponential}} function,
    \begin{equation}
    \begin{aligned}
                \Psi_{4,\c{\text{CHE}}}^{\c{(2,2)}} = &A_{\text{CHE}}\, \c{e^{-i\left[\omega_{\text{CHE}} \left(t - t^\text{CHE}_0\right) + \phi_\text{CHE}\right]}}\times\\
        &e^{-\left(t - t^\text{CHE}_0\right)^2 / 2\sigma_{\text{CHE}}^2}.
        \label{eq: singauss}
    \end{aligned}
    \end{equation}
    
    \item Second emission (merger and ringdown): divided into two components. The merger portion is described by an identical expression to Eq.~\eqref{eq: singauss} (with \c{``CHE''} labels replaced by ``M''), while the ringdown is modeled using \c{the first} fundamental quasinormal mode~\cite{Redondo-Yuste:2023ipg}, as follows:
    \begin{equation}
        \Psi^{\c{(2,2)}}_{4,\c{\text{R}}} = A_{\text{R}}\,
        \c{e^{-i\left[\omega_{\text{R}} (t - t_a) + \phi_{\text{R}}\right]}}
        e^{-(t - t_a)/\tau_{\text{R}}},
        \label{eq: ringdown}
    \end{equation}
    where $t_a$ marks the transition between the merger and ringdown \c{stages}. These decaying modes appear as the final BH radiates away its distortions, settling into a stable Kerr BH state.
\end{enumerate}

To ensure consistency and improve parameter accuracy, we first fit the amplitude envelope $\left|\Psi_4^{\c{(2,2)}}\right|$ using a simple Gaussian profile, obtaining \c{first the} values for $A_{\text{CHE}}$, $t_0^{\text{CHE}}$, and $\sigma_{\text{CHE}}$. \c{For this, the procedure we use is to minimize the residue} 
\begin{equation}
    \c{\sum_n \Bigl(\left|\Psi_{4,\, \rm CHE}^{(2,2)}\right|(t_n;A_{\text{CHE}},t_0^{\text{CHE}},\sigma_{\text{CHE}}) - \left|\Psi_{4,\, n}^{(2,2)}\right|\Bigr)^2},
\end{equation} 
\c{where $\left(t_n,\Psi_{4,n}^{(2,2)}\right)$ are our data points, and the free parameters are indicated after the semicolon.}

Considering these conditions, once the amplitude-related parameters are determined, we perform a fit of the \c{phase parameters: the angular frequency $\omega_{\rm CHE}$ and phase shift $\phi_{\rm CHE}$. For this, we minimize the following residue}
\begin{equation}
    \begin{aligned}
       \c{\sum_n} &\c{\Bigl|\Psi_{4,\, \rm CHE}^{(2,2)}\left(t_n,\boldsymbol{\theta}^{\rm CHE}_{\rm fit};\omega_{\rm CHE},\phi_{\rm CHE}\right) - \Psi_{4,n}^{(2,2)}\Bigr|^2},
  \end{aligned}
\end{equation}
\c{where $\boldsymbol{\theta}^{\rm CHE}_{\rm fit}=\left\{A_{\text{CHE}}^{\rm fit},t_{0,\rm fit}^{\text{CHE}},\sigma_{\text{CHE}}^{\rm fit}\right\}$. The same approach is applied to the merger \c{stage}.}

The ringdown portion of $\Psi_4^{\c{(2,2)}}$ is fitted in a similar fashion, but making use of the damped \c{complex exponential} phenomenological model in Eq.~\eqref{eq: ringdown}, instead of the Gaussian function. Since we restrict the analysis to the \c{fundamental quasi-normal} mode, the procedure remains tractable; including higher-order modes would require a more complex fitting scheme where the parameters of overlapping modes are not easily separable.

\begin{figure}[htbp]
    \centering
    \includegraphics[width=\columnwidth]{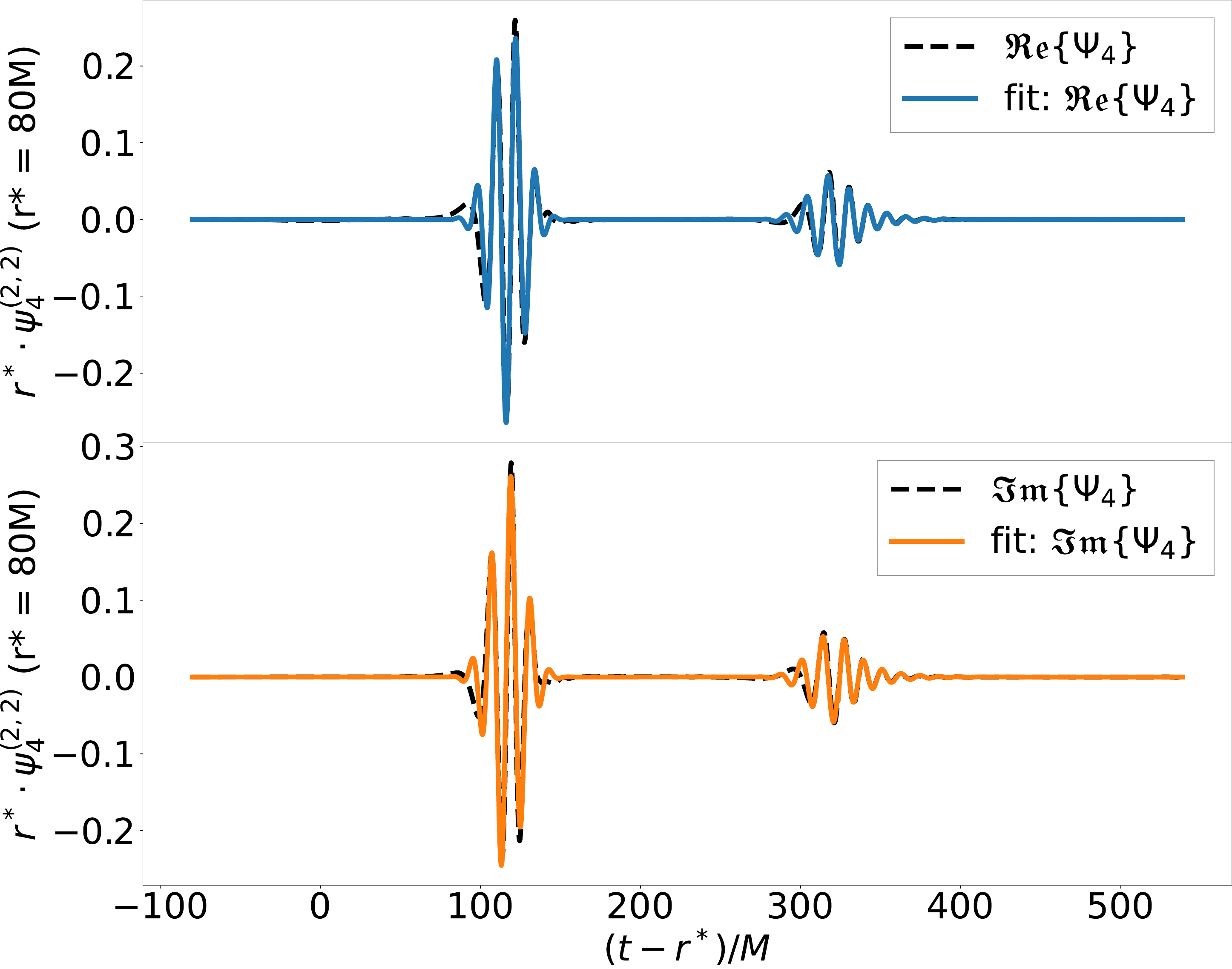}
    \caption{Real (blue) and imaginary (orange) \c{components} (upper and lower panels, respectively) \c{of the best-fit} compared to the simulated rescaled Weyl scalar (black dashed line) at a detector located at $r^*=80M$ from the c.m. The initial conditions correspond to the example case described in Sec.~\ref{subsec: gen. beh-DCs}.}
    \label{fig: fit-psi}
\end{figure}

In Fig.~\ref{fig: fit-psi}, we show an example of the fitting results for the same case depicted in Figs.~\ref{psi+tray} and~\ref{chi+mass}. The real and imaginary parts are shown separately, along with their fitted counterparts. The fits perform very well overall, particularly for the ringdown \c{stage}. The main discrepancies appear at the tails of the \c{Gaussian-modulated} fits, likely due to the gradual variation of the orbital frequency over time. This is consistent with Kepler’s law, where the gravitational-wave frequency obeys $\omega = 2\Omega$~\cite{10.1093/acprof:oso/9780198570745.001.0001}, and the orbital separation evolves during each emission.

\subsection{Time interval of gravitational-wave emissions}

In this and the following subsections, we determine effective trends for each of the fitted parameters of the Weyl scalar in terms of the incidence angle ($\theta$). These relations will later serve as the basis for constructing a phenomenological model of DC waveforms. Before analyzing these dependencies, however, we first carry out the \c{Gaussian-modulated} fits described previously and obtain a useful quantity that simplifies the discussion.

Using the parameters $t_0^{\text{CHE}}$ and $t_0^{\text{M}}$ [from Eq.~\eqref{eq: singauss}], we compute the temporal separation between the peaks of the two main emissions as
\begin{equation}
    \Delta t \equiv t_0^{\text{M}} - t_0^{\text{CHE}}.
\end{equation}
This quantity is physically meaningful: as the incidence angle increases, we approach the threshold value that separates pure CHEs (i.e., scattering events) from dynamical captures~\cite{Nelson:2019czq, Rettegno:2023ghr}. In the limit where the interaction becomes a pure CHE, the merger never occurs, implying $\Delta t \to \infty$. Therefore, we expect a divergence in the trend of $\Delta t$ as $\theta$ approaches this critical threshold angle $\theta_0$. This behavior is precisely what we observe in our simulations, as we show in the following.

In addition to determining the CHE/DC transition, another useful feature of $\theta_0$ is that it allows us to ``normalize'' our horizontal axis when obtaining the trends of each parameter, leading to more compact and consistent visual representations despite the wide range of $\theta$ (and $b/M$) values explored in Table~\ref{Tab: Cases}.

\begin{figure}[htbp]
    \centering
    \includegraphics[width=\columnwidth]{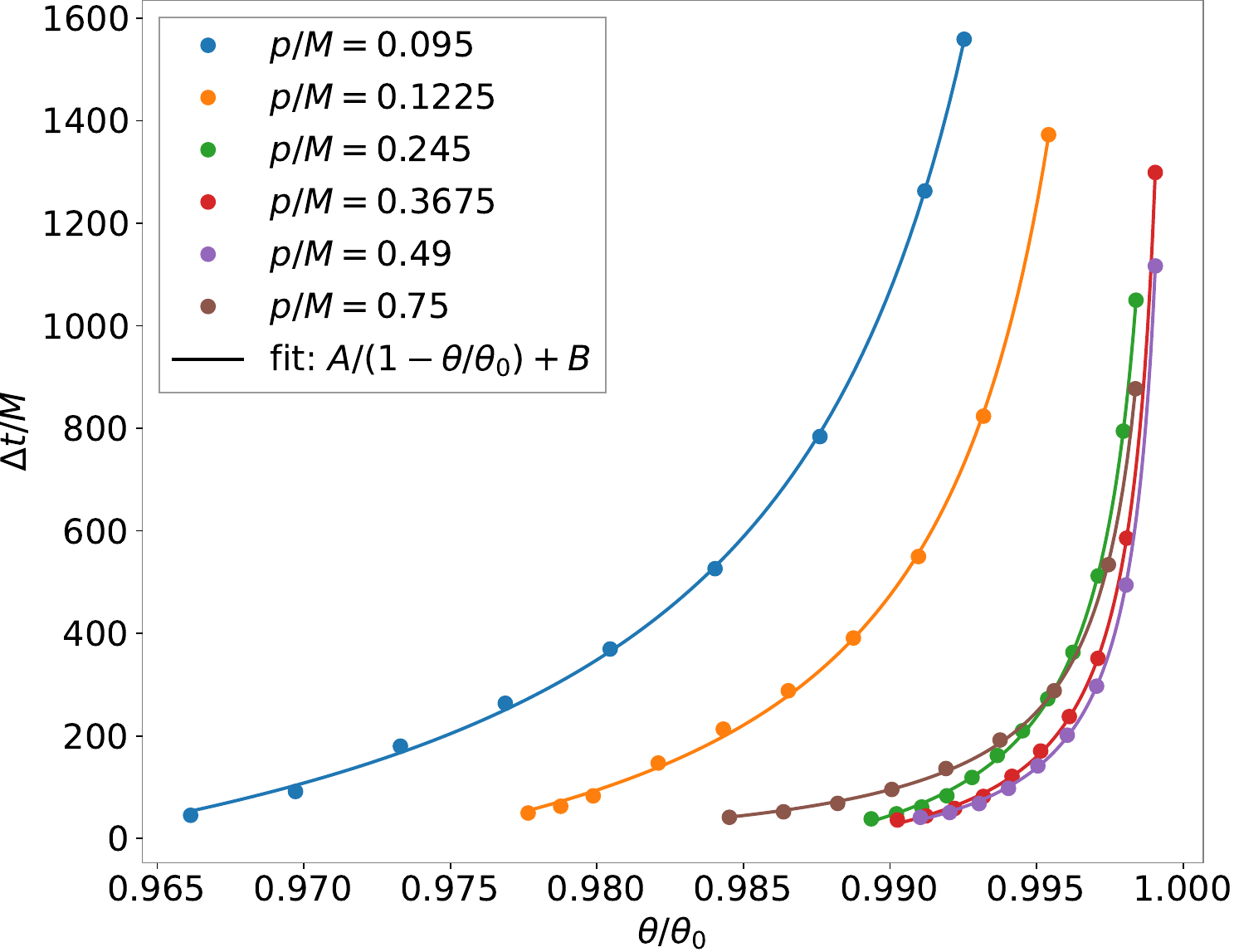}
    \caption{Time intervals between peaks, $\Delta t = t_0^{\text{M}} - t_0^{\text{CHE}}$, as a function of the normalized incidence angle $\theta/\theta_0$. The data points are obtained from the \c{Gaussian-modulated} fits, while the solid curves (which diverge at $\theta/\theta_0=1$) represent fits using Eq.~\eqref{eq: divergence}.}
    \label{fig: times-ang}
\end{figure}

Figure~\ref{fig: times-ang} shows $\Delta t/M$ for each scenario, together with the best-fit function
\begin{equation}
    \Delta t/M = \frac{A}{1 - \theta/\theta_0} + B,
    \label{eq: divergence}
\end{equation}
whose parameters ($A$, $B$, and $\theta_0$) are listed in Table~\ref{Tab: fit-times}. \c{In this figure, we can also see why the incidence angle ranges in Table~\ref{Tab: Cases} are broader for smaller momenta: the time between both emissions decays more slowly as the angle decreases, which allows us to resolve both emission peaks for a wider angular range. As we consider an event as a pure merger when both peaks start to mix, this transition occurs for smaller $\theta/\theta_0$ in the scenarios with lowest momenta.}

\begin{table}[htbp]  
\caption{Best-fit parameters obtained from Eq.~\eqref{eq: divergence}. Uncertainties in parentheses correspond to 68\% CL errors on the last digit(s). The fitted curves are shown alongside the numerical data in Fig.~\ref{fig: times-ang}.}
\centering 
\renewcommand{\arraystretch}{1.5}
\setlength{\tabcolsep}{3.3pt}
\begin{tabular}{|c|c|c|c|c|}
\hline  
 $p/M$ & $\theta_0$ (deg) &$A$ & $B$ \\ [0.5ex]  
\hline   
$0.095$  &$ 6.404(1)$ & $14.\c{4}(6)$ & $-3\c{72(19)}$ \\ [0.5ex] 
\hline   
$0.1225$  &$ 5.1629(7)$ & $7.6(2)$ & $-285(14)$ \\ [0.5ex] 
\hline   
$0.245$  &$ 3.3299(1)$ & $1.92(4)$ & $-147(6)$ \\ [0.5ex]
\hline
$0.3675$  &$2.93060(4)$ & $1.33(2)$ & $-108(3)$ \\ [0.5ex]
\hline
$0.49$  &$2.85889(4)$ & $1.14(2)$ & $-91(3)$ \\ [0.5ex]
\hline 
$0.75$  &$3.0997(1)$ & $1.51(3)$ & $-56(4)$ \\
\hline
\end{tabular}  
\label{Tab: fit-times}
\end{table} 

\begin{figure}[htbp]
    \centering
    \includegraphics[width=\columnwidth]{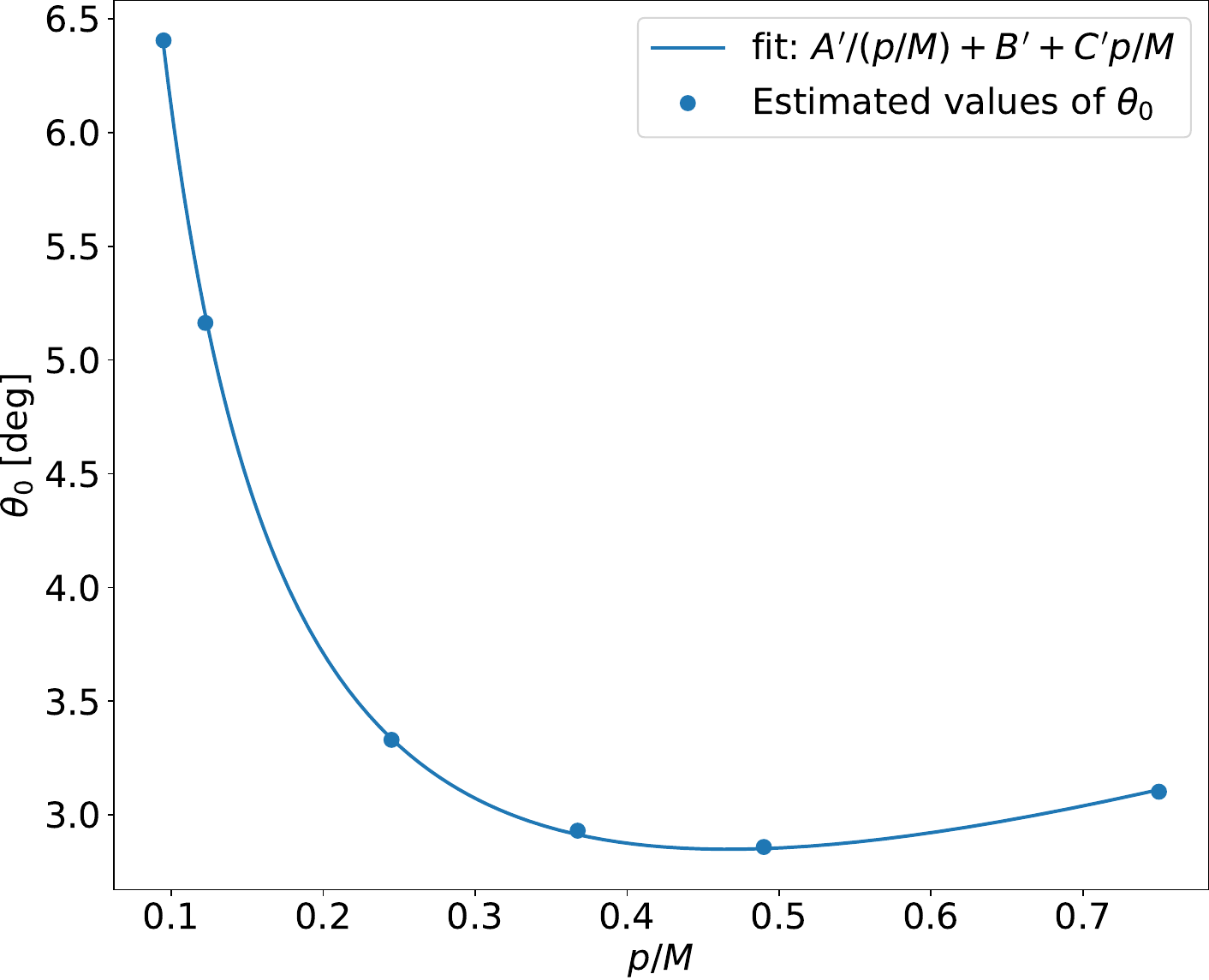}
    \caption{Estimated threshold angles $\theta_0$ (from Table~\ref{Tab: fit-times}) as a function of the initial momentum $p/M$. The solid curve represents the best-fit model given by Eq.~\eqref{eq: theta0 vs p}.}
    \label{fig: ang0}
\end{figure}

With this, we can now analyze the trend of $\theta_0$ as a function of the initial momentum $p/M$. As discussed earlier (see Sec.~\ref{subsec: gen. beh-DCs}), higher momenta correspond to smaller threshold angles: the faster a black hole moves, the closer it must pass to the other to achieve capture. Interestingly, this monotonic behavior breaks down for the most relativistic cases ($p/M \gtrsim 0.49$), where the threshold angle increases again. The extracted $\theta_0$ values from Table~\ref{Tab: fit-times} are plotted in Fig.~\ref{fig: ang0}.

The best-fit model describing this relation is
\begin{equation}
    \theta_0(p/M) = \frac{A'}{p/M} + B' + C'p/M,
    \label{eq: theta0 vs p}
\end{equation}
with coefficients $A' = 0.529(6)$, $B' = 0.58(6)$, and $C' = 2.43(9)$ [deg] (1$\sigma$ uncertainties shown in parentheses).

The observed upturn in $\theta_0$ at large $p/M$ remains physically unclear. A plausible explanation involves enhanced spin-orbit and spin-spin couplings at high velocities, since spin induction becomes significant for $p/M > 0.3675$. Following Ref.~\cite{Rodr_guez_Monteverde_2025}, the leading-order spin-spin Hamiltonian can be expressed as
\begin{equation} 
H_{S_1S_2}\simeq-\frac{G}{c^2r^3}\Big(\vec{S}_1\cdot\vec{S}_2 - 3(\vec{S}_1\cdot\hat{\vec{r}}_1)(\vec{S}_2\cdot\hat{\vec{r}}_2)\Big),
\label{eq: spin-spin} 
\end{equation}
where $\vec{S}_j$ and $\hat{\vec{r}}_j$ denote the spin and unit position vector of each black hole ($j=1,2$), respectively. In this framework, parallel spins contribute negatively to the Hamiltonian, effectively increasing the attraction between the black holes. Thus, significant spin induction aligned with the orbital angular momentum ($\vec{L}$) facilitates the capture event, increasing $\theta_0$ for larger $p/M$. \c{A different argument for this behaviour is given in~\cite{Nelson:2019czq}, where the authors argue that incidence angles corresponding to different scenarios ($p/M$) are not immediately comparable because their initial dimensionless separation in terms of the total ADM mass of the system, $d/M_{\rm ADM}$, are different, decreasing with higher $p/M$.}

\begin{figure*}[htbp]
    \centering
    \includegraphics[width=\textwidth]{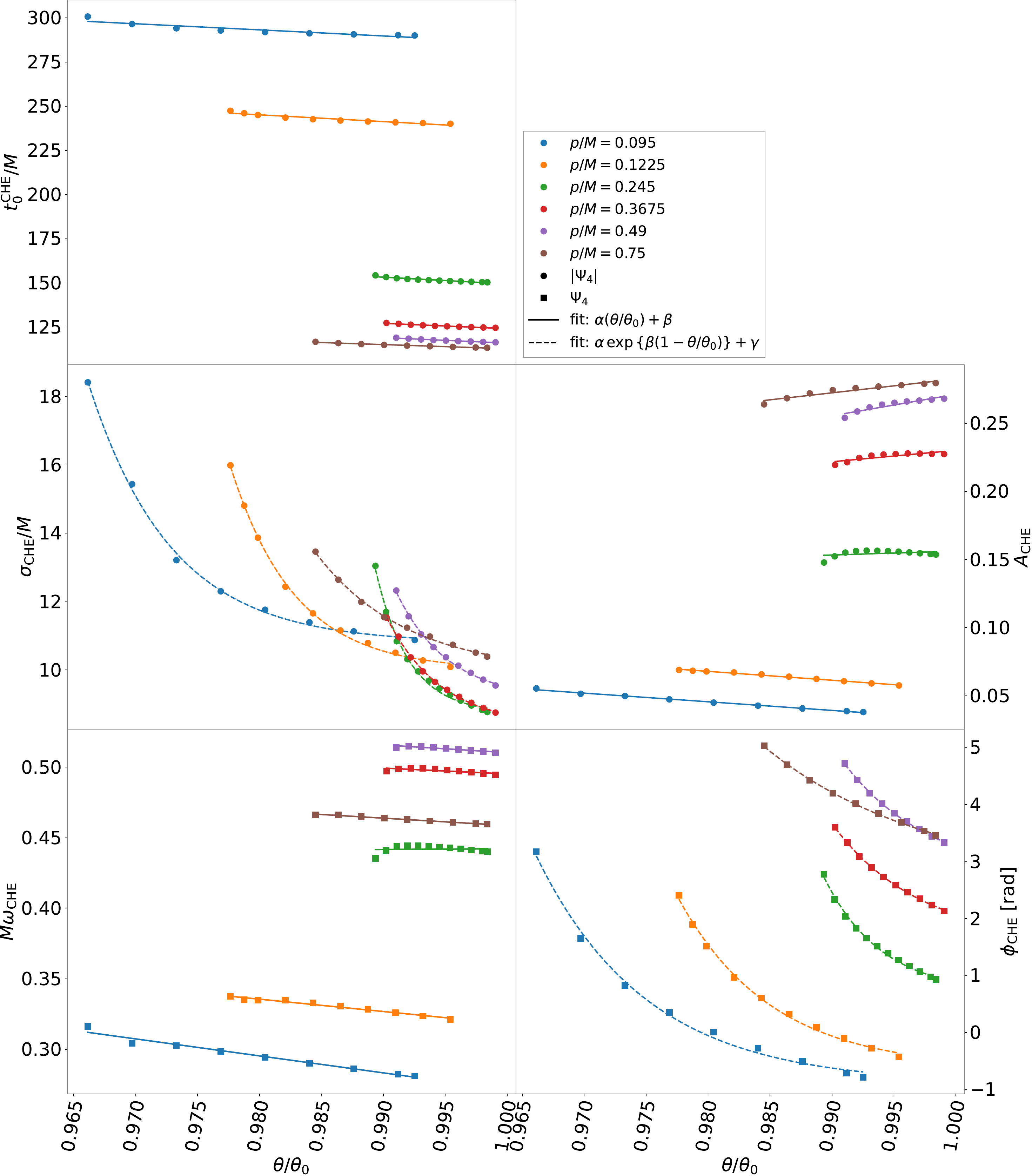}
    \caption{Estimated parameters [obtained using Eq.~\eqref{eq: singauss}] for the CHE portions of the Weyl scalars for all $p/M$ values. We fit these to linear fits (solid lines) which follow Eq.~\eqref{eq: linear trend}, and to exponential decays (dashed lines) which follow Eq.~\eqref{eq: expo. decay}. Solid circles denote values obtained from the amplitude fit, while solid squares correspond to the sinusoid fit. The full set of fitted coefficients and their uncertainties is provided in Table~\ref{Tab: Annex}. The values of the phase shifts have been extended to the range $(-2\pi,2\pi]$ to properly study the observed trends.}
    \label{fig: CHE-params}
\end{figure*}

\subsection{Analysis and model of the CHE emission}
\label{Subsec: CHE}

We now focus on the fitted parameters describing the first (CHE) emission, as illustrated in Fig.~\ref{fig: fit-psi}. Depending on the observed behavior, we model these parameters using either a linear relation,
\begin{equation}
    y = \alpha(\theta/\theta_0) + \beta,
    \label{eq: linear trend}
\end{equation}
or an exponential decay,
\begin{equation}
    y' = \alpha\,\exp\{\beta(1 - \theta/\theta_0)\} + \gamma,
    \label{eq: expo. decay}
\end{equation}
where $y \in \{A_{\rm CHE},\ \omega_{\rm CHE},\ t_0^{\rm CHE}\}$ and $y' \in \{\sigma_{\rm CHE},\ \phi_{\rm CHE,R},\ \phi_{\rm CHE,I}\}$.

As shown in Fig.~\ref{fig: CHE-params}, the \c{Gaussian-modulated} model describes the CHE emission with excellent accuracy, and the extracted parameters exhibit smooth and physically consistent trends. Two of them ($A_{\text{CHE}}$ and $\omega_{\text{CHE}}$) show slight deviations from strict linearity, suggesting mild nonlinear dependencies that may carry physical meaning. Both parameters increase with $\theta/\theta_0$ up to a critical point and then change slope, likely reflecting that weaker GW bursts occur for larger incidence angles, while nearly head-on collisions also require less radiation to achieve capture.

The coefficients and uncertainties obtained from these fits are reported in Table~\ref{Tab: Annex}, in the Appendix~\ref{app: pars_merger_ringdown}. Overall, this modeling strategy provides a robust and physically motivated description of the CHE waveform. Future refinements could address small discrepancies in the signal tails by considering time-dependent frequencies or expansions to higher-order modes.

Finally, we note that given the current sensitivity of the LVK detector network, the resulting phenomenological waveform model may already suffice for parameter estimation or signal detection of CHE events, which remain observationally elusive.

\subsection{Analysis and model of the merger-ringdown emission}
\label{Subsec: merger, ringdown}

The final emission (the merger and ringdown) is modeled as two contiguous \c{stages}, ensuring continuity and differentiability between Eqs.~\eqref{eq: singauss} and~\eqref{eq: ringdown} at the given transition time ($t_a$). This time is chosen as $t_a = t_0 + 5M$, where $t_0$ corresponds to the maximum of the second emission (typically close to $t_0^{\rm M}$).
\subsubsection{Merger \c{stage}}
The estimated parameters for the merger portion are summarized in Appendix~\ref{app: pars_merger_ringdown} (see Fig.~\ref{fig: Merger-params}). Unlike the CHE case, no clear functional trends with $\theta/\theta_0$ were found, so we did not perform further fitting of the estimated parameters as was indeed done for the first emission in the previous subsection. There are three main factors which could explain this phenomenon:
\begin{enumerate}[i.]
    \item The estimated parameters for all scenarios (values of $p/M$) are of similar order of magnitude, leading to tightly clustered data points. This indicates that once a bound system forms after the CHE, the merger emits comparable amounts of energy and angular momentum regardless of the initial conditions.
    \item The division between merger and ringdown is somewhat arbitrary. The start of the ringdown occurs at slightly different times for the real and imaginary components, making the use of a single cutoff time ($t_a$) an approximation. In fact, there are entire works that tackle this concrete issue~\cite{Bhagwat_2018}, and may be reason for further work. 
    \item We use the entire merger-ringdown emission to estimate the parameters of the merger \c{stage}, which may be blurring the boundary between the two regimes. This choice was made because we checked that restricting the fit to only the rising portion of the peak worsens the results due to the strong nonlinear dynamics present at this stage (and, especially as well, due to the small number of sample points to estimate the necessary parameters).
\end{enumerate}

Despite these issues, the fits perform reasonably well overall. The amplitude fit worsens particularly when both emissions (CHE and merger-ringdown) overlap significantly ($\Delta t/M \lesssim 100$), since the signals interfere nonlinearly at their tails (see an example of this in Fig.~\ref{fig: fit-psi-2} in Appendix~\ref{app: overlap}). Conversely, the $t_0^{\text{M}}$ parameter shows excellent agreement, as the fitted $\Delta t + t_0^{\text{CHE}}$ values reproduce the simulated merger times remarkably well.
\subsubsection{Ringdown \c{stage}}
The fits for the ringdown portion (see Fig.~\ref{fig: Ring1-params} in Appendix~\ref{app: pars_merger_ringdown}) follow the same methodology as before: we first fit the amplitude, fix its value, and then fit the \c{phase} [using Eq.~\eqref{eq: ringdown}] \c{to find the frequency and phase shift parameters}. Although the fits reproduce the time evolution of the Weyl scalar quite accurately (as shown in Fig.~\ref{fig: fit-psi}), the fitted parameters themselves show mild oscillations and no analytical trends, particularly for smaller values of $\theta/\theta_0$. 

Future improvements will involve refining the cutoff time ($t_a$) that marks the start of the ringdown, potentially delaying it further to isolate the pure quasinormal mode regime or applying techniques used in the literature~\cite{Bhagwat_2018}, and improving the fits of the merger portion simultaneously. 

While the current description already captures precise magnitudes of the relevant parameters, a more accurate treatment of the transition between merger and ringdown \c{stages} will be crucial for waveform modeling and GW data analysis applications.

\section{Black hole properties}
\label{sec: Num. res. BH}
After analyzing the properties of the measured Weyl scalar, we now turn to the study of the intrinsic properties of BHs: their spins and masses, as well as additional quantities derived from them. These quantities are computed using the \texttt{QuasiLocalMeasures} thorn of the Einstein Toolkit, as detailed in Sec.~\ref{subsec: spin measurements}.

All variables describing the final state of the BH are evaluated once the system reaches a relaxed configuration. For all cases, this time was chosen to be $140M$ after the merger.

For this section, we exclude the data corresponding to the most relativistic configuration, $p/M=0.75$. Although the associated waveforms are reliable, the final BHs obtained in these simulations present significant numerical uncertainties that prevent us from extracting physically meaningful results for their final states. \c{The origin of these uncertainties is the numerical resolution, as for $p/M=0.75$ \texttt{AHFinderDirect} fails to resolve the merged black hole, producing unphysical results for its mass and spin.}

\subsection{Analysis of spin}

\begin{figure*}[htbp]
    \centering
    \includegraphics[width=\textwidth]{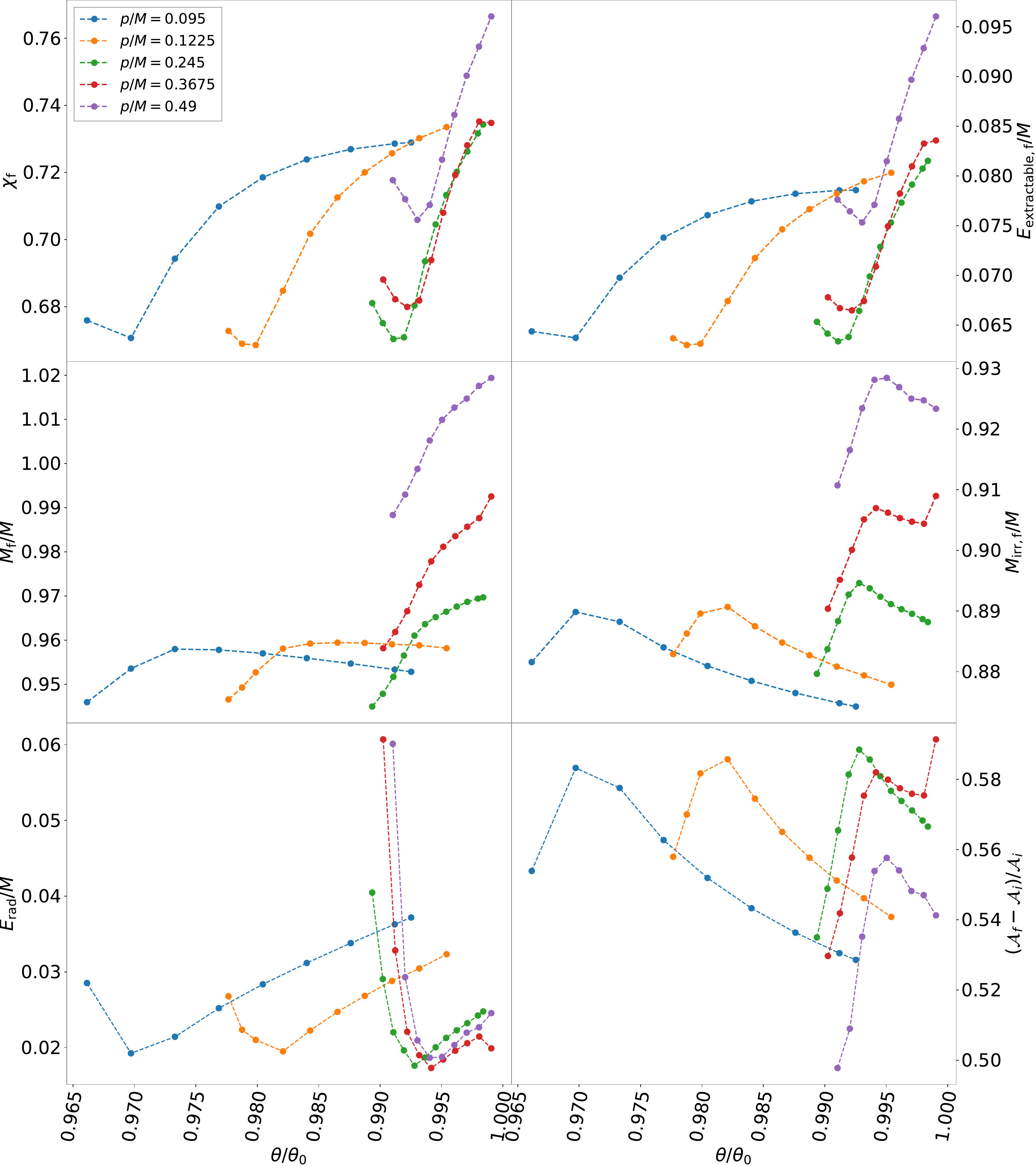}
    \caption{Spin (upper left panel), extractable energy (upper right panel), mass (middle left panel) and irreducible mass (middle right panel) of the final BH as functions of the normalized angle $\theta/\theta_0$ for each scenario. With respect to the same variable in the horizontal axis, we show the total radiated \c{energy} (lower left panel) when the BHs merge and the fractional area difference (lower right panel) between the final BH and the ones before the merger. Final values have been computed at a time of $140M$ after the merger, while values prior to the merger (and after the CHE) have been calculated at a time $30M$ before the coalescence.}
    \label{fig: fin-ext-spins}
\end{figure*}

In Fig.~\ref{fig: fin-ext-spins} (upper left panel), we show the spin of the remnant BH measured at $140M$ after the merger for each case. The final spin lies within the range $\chi_{\rm f}\in[0.67,0.77]$. We observe that the spin initially decreases with increasing $\theta/\theta_0$, reaches a minimum (which varies slightly for each $p/M$), and then increases again. This nonmonotonic behavior can be interpreted similarly to the small deviations found in the parameters ($A_{\text{CHE}},\omega_{\text{CHE}}$) discussed in Sec.~\ref{Subsec: CHE}: for small $\theta/\theta_0$, corresponding to near head-on collisions, the GW emission during the CHE is weak (enough to bind the orbit but not significant enough to cause major energy or angular momentum losses). As $\theta/\theta_0$ increases, the emission becomes stronger, maximizing the energy and angular momentum loss and hence reducing the final spin. For even larger angles, the encounters become weaker again, leading to less GW emission and a corresponding increase in the final spin, as more angular momentum remains available to be converted into the spin of the remnant BH.

The rightmost red point in Fig.~\ref{fig: fin-ext-spins} deviates from the overall increasing trend, but this is likely due to numerical inaccuracies. Simulations at such large angles require longer evolution times, which accumulate significant numerical errors. This outlier will also affect the mass analysis later on.

It is also noticeable that the two lowest values of $p/M$ yield nearly identical trends in their final spins. The green and red curves exhibit similar magnitudes to the blue and orange ones but develop a clearer positive slope beyond the critical point. The purple curve ($p/M=0.49$) differs most from the others, likely because higher $p/M$ values induce stronger spins during the CHE \c{stage}, as reported in Refs.~\cite{Jaraba:2021ces, Nelson:2019czq}. For $p/M=0.095-0.245$, the merger occurs with negligible induced spin, leaving mainly the remnant orbital angular momentum to contribute to $\chi_{\rm f}$. In contrast, the $p/M=0.49$ case exhibits substantial induced spins ($\chi\sim 0.14$) before the merger, leading to a higher final spin. The $p/M=0.3675$ case lies between these extremes, with induced spins of order $\chi\sim 0.07$.

\subsection{Analysis of the BH masses}

In Sec.~\ref{Subsec. app-hor}, we established the relation between ADM and irreducible masses [see Eq.~\eqref{eq: ADM-irreducible}]. These quantities are displayed in Fig.~\ref{fig: fin-ext-spins}, where the middle left and middle right panels show $M_{\rm f}$ and $M_{\rm f,irr}$, respectively.

In the upper right panel of Fig.~\ref{fig: fin-ext-spins}, we plot the extractable energy (defined as $E_{\text{extractable}}\equiv M_{\rm f}-M_{\rm f,\text{irr}}$) which represents the maximum amount of energy that can be extracted from a Kerr BH through processes such as the Penrose mechanism or BH mergers~\cite{ruffini2024roleirreduciblemassrepetitive}. In the upper panels of this figure, one can see that the trends of $E_{\text{extractable}}$ closely follow those of $\chi_{\rm f}$, suggesting a strong direct dependence between both quantities. This can be explained by expanding the irreducible mass formula,
\begin{equation}
    M_{\rm irr,f}=M_{\rm f}\sqrt{\frac{1+\sqrt{1-\chi_{\rm f}^2}}{2}},
\end{equation}
around $\chi_{\rm f}=0.686$, for reference (the typical final spin for quasicircular mergers). With this, we obtain
\begin{equation}
    \frac{E_{\rm extractable,f}}{M_{\rm f}}\approx 0.071+0.235(\chi_{\rm f}-0.686),
\end{equation}
which explains the almost linear correlation observed between the extractable energy and the final spin, as our results remain close to $\chi_{\rm f}=0.686$, making the linear approximation valid.

The extractable energy varies within $E_{\text{extractable}}\in[0.064,0.097]$, with the largest values corresponding to the most relativistic encounters ($p/M=0.49$), which exhibit the highest induced spins and therefore the greatest potential for energy extraction.

Additionally, in the middle left panel of Fig.~\ref{fig: fin-ext-spins}, we observe that the final mass generally increases with $\theta/\theta_0$, particularly for $p/M=0.3675$ and $p/M=0.49$. For smaller $p/M$ values, the slope changes sign near the critical point: the orange curve flattens, while the blue one turns slightly negative. The origin of this transition is not entirely clear; it likely reflects the interplay between GW emission, spin induction, and mass increase. 

Future work should focus on quantifying the correlations between spin, ADM mass, and irreducible mass (which consistently show a negative slope beyond the critical point) to better understand the underlying mechanisms. Effects such as angular momentum redistribution, tidal torquing, and tidal heating through horizon absorption\footnote{Horizon absorption can explain why $M_{\rm f}$ exceeds $1M$ in some cases (see the middle left panel of Fig.~\ref{fig: fin-ext-spins}), since after the CHE the individual masses can reach $m/M\sim 0.52$, thus increasing the total ADM mass.} all likely contribute to the complex trends observed in these simulations of dynamical captures~\cite{Chiaramello:2024unv, PhysRevD.87.044022}. In future works, we also plan to compare the obtained final spins and masses with those predicted for quasicircular and head-on mergers with equivalent effective spins, expecting our results to lie between these limiting cases.

\subsection{Analysis of radiated mass and BH areas} 

Finally, we evaluate the radiated \c{energy} during the merger. When two BHs coalesce, they emit large amounts of energy and angular momentum as gravitational waves. The remaining angular momentum contributes to the final spin, while \cb{the remaining energy defines the final BH energy}. Consequently, the \cb{spin and energy} of the final BH \cb{are} smaller than the \cb{total angular momentum and energy} of the progenitors, \cb{respectively}. 

\c{Using GW energy flux at a distance $r^*=80M$ we calculate the cumulative energy loss of the system during the merger to compute how much energy was radiated away during the formation of the final BH. Although the energy flux is evaluated at a finite extraction radius $r^*=80M$, this distance is sufficiently large that near-zone effects are negligible and the waveform is already in the radiation zone.}

The results are shown in the lower left panel of Fig.~\ref{fig: fin-ext-spins}, where we find that more head-on collisions emit more \c{energy} as GW radiation, and that more relativistic encounters radiate a larger fraction of the total \cb{energy}. These mergers radiate roughly between $\c{1.8}\%$ and $\c{6}\%$ of the total initial \cb{energy} as GWs; a range consistent with typical LVK observations (for instance, GW150914 radiated about $4.6\%$ of its total mass~\cite{Abbott_2016,Abbott_2023}).

Future detections may reveal events with even higher radiated \c{energy} percentages ($8\%-10\%$), potentially signaling highly relativistic BH encounters ($p/M\gg0.49$) on nearly head-on trajectories; a distinctive signature of dynamical capture. \c{In the limiting case of head-on trajectories, all energy is emitted in a single burst instead of two, so the radiated energy is expected to be significantly greater than the values obtained throughout this study, which we will explore in future work.}

Lastly, we analyze the change in horizon area. According to Hawking’s area theorem, the total BH area cannot decrease over time through classical processes, implying $\mathcal{A}_f>\mathcal{A}_i=\mathcal{A}_{1,i}+\mathcal{A}_{2,i}$. This was recently confirmed using ringdown measurements~\cite{KAGRA:2025oiz}. Using the \texttt{QuasiLocalMeasures} thorn, we can directly compute these areas before and after merger. Figure~\ref{fig: fin-ext-spins} shows that all our simulations satisfy this condition, in full agreement with Hawking’s area law.

\section{Conclusions}
\label{sec: Conclusions}
In this paper, we have presented a first comprehensive approach to the analysis of dynamical captures that culminate in a merger. Our primary objective was to model the outgoing gravitational radiation of such events and to study the trends of the spins, masses, and irreducible masses of the final black holes as functions of the incidence angle.  

Using the waveform-fitting procedure described throughout this work, we obtained an accurate parametrization of the time separation between the close hyperbolic encounter and the merger-ringdown emissions ($\Delta t/M$), as shown in Eq.~\eqref{fig: times-ang}. This characterization allowed us to define the threshold angle $\theta_0$, which marks the transition between purely scattering encounters and dynamical captures.  

Furthermore, as discussed in Sec.~\ref{Subsec: CHE}, the \c{Gaussian-modulated} parametrization introduced in Eq.~\eqref{eq: singauss} reproduces the Weyl scalar with remarkable precision. The fitted parameters exhibit clear and consistent trends with respect to the normalized incidence angle $\theta/\theta_0$, indicating that this model captures the essential physical features of the CHE emission \c{stage}.  

In contrast, the parameters describing the merger and ringdown emissions display less regular behavior, as seen in Appendix~\ref{app: pars_merger_ringdown}. Some of these quantities, particularly the phase shifts, show mild oscillations or clustering around similar orders of magnitude, complicating the identification of global trends using analytical functions. Nevertheless, the model described in Sec.~\ref{Subsec: merger, ringdown} [Eqs.~\eqref{eq: singauss} and~\eqref{eq: ringdown}] achieves good overall agreement with the NR Weyl scalar (see Fig.~\ref{fig: fit-psi}), providing a robust phenomenological description of the characteristic frequencies, amplitudes, and timescales that shape the final emission.  

Additionally, Fig.~\ref{fig: fin-ext-spins} illustrates the dependencies of the final black hole spin, extractable energy, ADM mass, irreducible mass, radiated energy, and fractional area difference on $\theta/\theta_0$ across different initial conditions. These results highlight the complex interplay between gravitational-wave emission, initial angular momentum, and relativistic mechanisms such as spin induction~\cite{Jaraba:2021ces,Rodr_guez_Monteverde_2025, Nelson:2019czq} and horizon absorption (tidal heating and torquing)~\cite{Chiaramello:2024unv, PhysRevD.108.044039}. A deeper understanding of these effects will be crucial to fully interpret the observed behaviors and to refine future models of dynamical-capture phenomenology.  

In conclusion, this paper lays the groundwork for the development of a comprehensive waveform model for dynamical captures. Such a model will not only improve our understanding of the waveform morphology, but also provide insights into the underlying orbital dynamics, spin interactions, and mass-energy exchange mechanisms that govern these extreme events. Future work will focus on extending and refining this framework, ultimately leading to a more complete picture of CHEs and DCs, and perhaps also studying whether these events occur as well for neutron stars~\cite{fontbuté2025gravitationalscatteringneutronstars}. As detector sensitivities continue to improve, these results are expected to play an important role in the potential detection and interpretation of GW signals from dynamical capture events.

\begin{acknowledgments}
    All the simulations have been run in the Hydra HPC cluster at the Instituto de F\'isica Te\'orica (IFT). J.G.B. acknowledges support from the Spanish Research Project No. PID2024-159420NB-C43 [MICINN-FEDER], and the Centro de Excelencia Severo Ochoa Program No. CEX2020-001007-S at IFT. S.J. acknowledges support from the Agence Nationale de la Recherche (ANR) under Contract No. ANR-22-CE31-0001-01.
\end{acknowledgments}

\section*{Data availability}

The data that support the findings of this article are not publicly available. The data are available from the authors upon reasonable request.

\appendix

\section{Impact parameters estimated from the black hole trajectories}
\label{app: impact pars}

After the initial conditions stabilize, and before the black holes interact, we can fit their trajectories to hyperbolic-like curves. We choose to fit the part of the trajectory corresponding to the times $30-38M$. From these hyperbolae, one can get the value of the impact parameter of each case. 

\begin{figure}[htbp]
    \centering
    \includegraphics[width=\columnwidth]{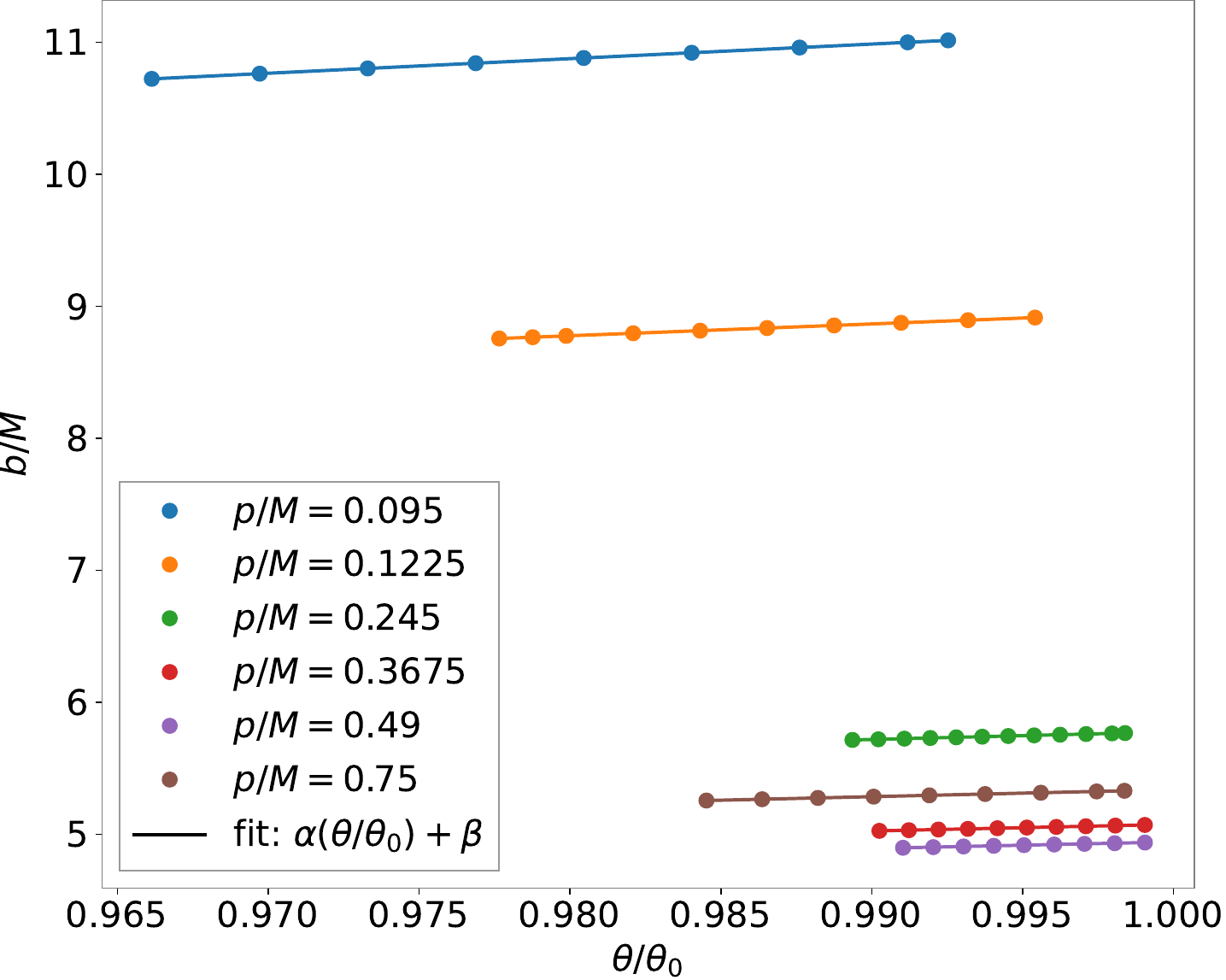}
    \caption{Estimated impact parameters ($b/M$) in terms of the normalized incidence angle $\theta/\theta_0$ for each scenario.}
    \label{fig: impact-params}
\end{figure}

\begin{table}[htbp]
\caption{Best-fit parameters obtained from Eq.~\eqref{eq: linear trend} for the estimated impact parameters. Uncertainties in parentheses correspond to 68\% CL errors on the last digit(s). The fitted curves are shown alongside the numerical data in Fig.~\ref{fig: impact-params}.}
\centering 
\renewcommand{\arraystretch}{1.5}
\setlength{\tabcolsep}{3.3pt}
\begin{tabular}{|c|c|c|c|}
\hline  
 $p/M$ & $\alpha$ &$\beta$  \\ [0.5ex]  
\hline   
$0.095$  &$ 11.05\c{57}(2)$ & $0.042\c{4}(2)$ \\ [0.5ex] 
\hline   
$0.1225$  &$ 8.9343(4)$ & $0.0221(4)$ \\ [0.5ex] 
\hline   
$0.245$  &$ 5.773(2)$ & $0.003(1)$ \\ [0.5ex]
\hline
$0.3675$  &$5.0740(5)$ & $0.0021(4)$ \\ [0.5ex]
\hline
$0.49$  &$4.9408(5)$ & $0.0018(5)$ \\ [0.5ex]
\hline 
$0.75$  &$5.3358(6)$ & $0.0029(5)$ \\
\hline
\end{tabular}  
\label{Tab: impact-params}
\end{table} 

\begin{figure*}[htbp]
    \centering
    \includegraphics[width=\linewidth]{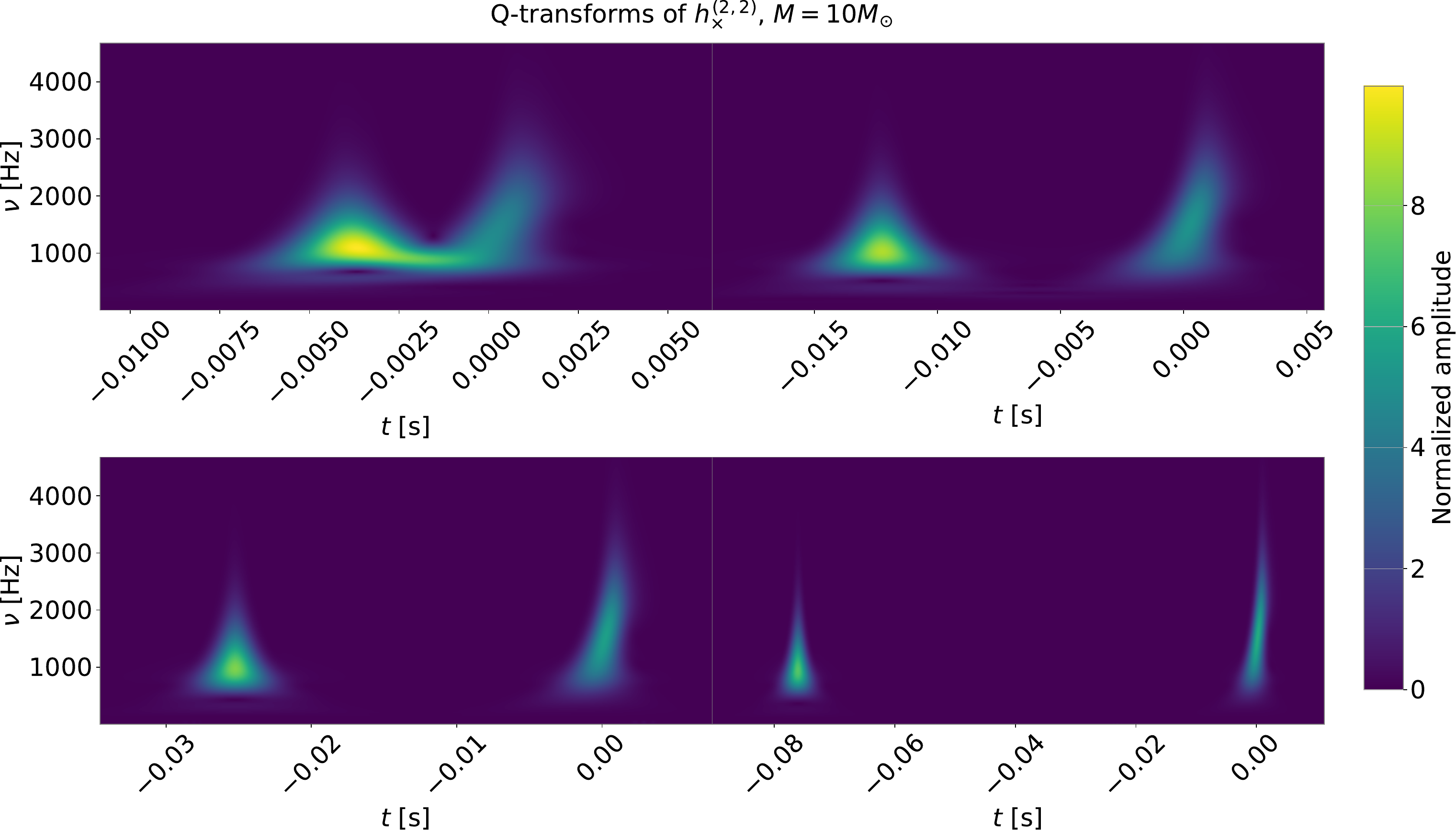}
    \caption{Example of Q-transforms over the dominant mode of the cross polarization ($h_\times$) of the strain from NR simulations with $p/M=0.095$. The angles of incidence (from left to right and from top to bottom) are: $\theta=6.21,\, 6.26,\, 6.30, \, 6.36^\circ$. The chosen initial total mass is $M=10M_\odot$. For these spectrograms, we mark $t=0$ as the merger time.}
    \label{fig: Q-plots}
    \end{figure*}

Since we are in a regime of small angles (see Table~\ref{Tab: Cases}), we expect that the trend followed by the impact parameters with respect to the incidence angle is linear. This can be seen in Fig.~\ref{fig: impact-params}, where we have included a linear fit [Eq.~\eqref{eq: linear trend}] for the estimated impact parameters.

We can check that indeed, in the fits of the estimated impact parameters, $\beta\approx 0$ (see Table~\ref{Tab: impact-params}), which implies that we recover the small-angle approximation for the impact parameter, $b/M\approx \alpha\theta/\theta_0$, showing that DCs occur within this regime of small incidence angles (or impact parameters).

In fact, we have can easily see that $\alpha/\theta_0\approx 100$, as one would expect from the setup described in Sec.~\ref{Subsec: setup}. However, the analogous value of the threshold incidence angle for the impact parameter is properly obtained (within this framework) as $b_0/M=\alpha+\beta$.

\section{Examples of Q-transforms}
\label{app: Q-plots}

In the context of GW detection, it is common to visualize GW signals through spectrograms, also called Q-plots~\cite{Abbott_2016, Abbott_2023}. Therefore, as visual illustrations of these for DCs, in Fig.~\ref{fig: Q-plots} we show four examples (using an initial total mass of $10M_\odot$) of Q-plots for the cross polarization of the simulated \c{$l=m=2$ mode of the} strain, $h_\times^{\c{(2,2)}}$. We see that larger incidence angles separate the peaks, as expected from our analysis. Additionally, it can be noted that when increasing the incidence angle, the intensity of the second emission starts growing, while the one of the first emission diminishes. 

\begin{figure*}[htbp]
    \centering
    \includegraphics[width=\linewidth]{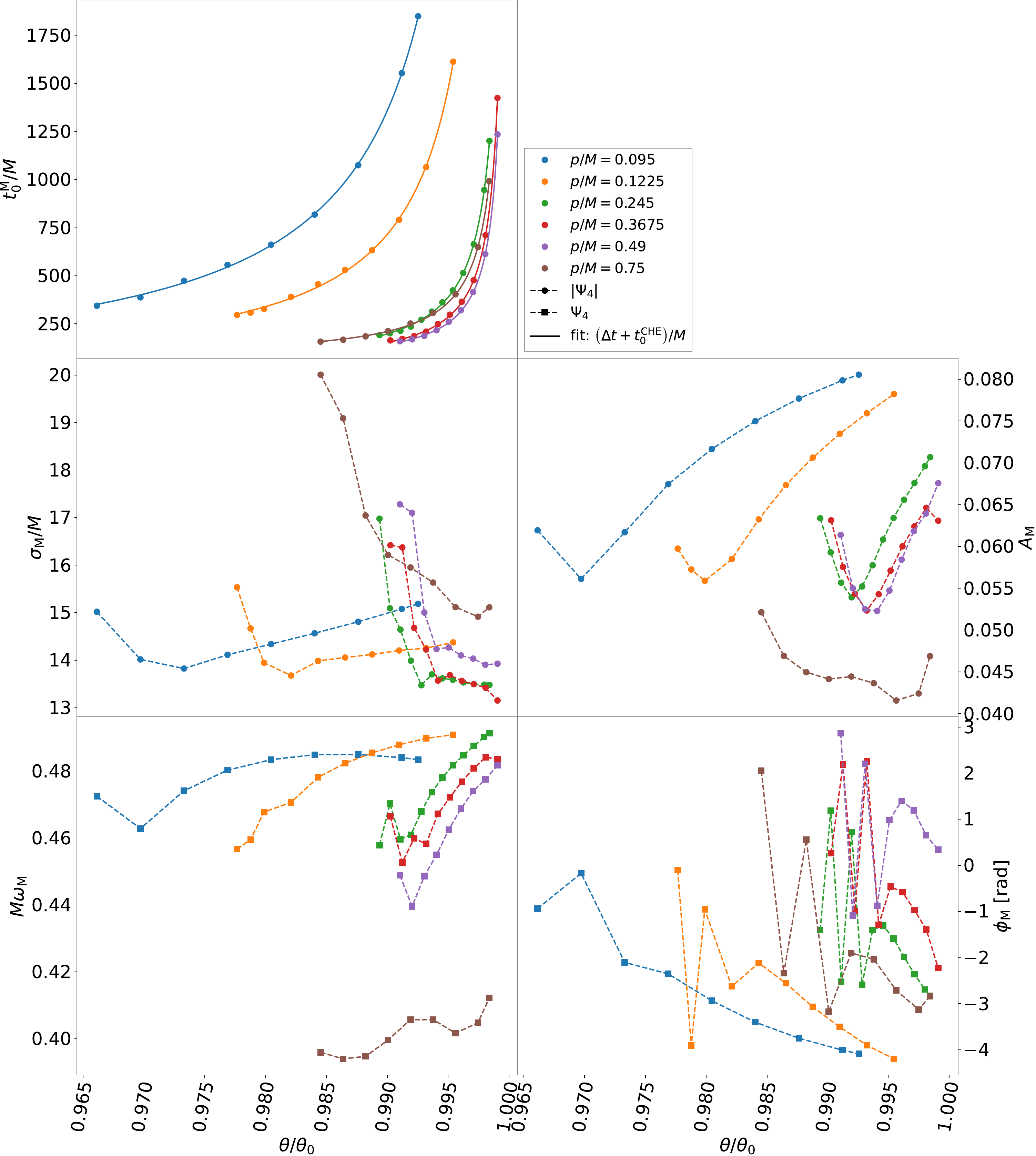}
    \caption{Estimated parameters from the fitted merger portions of the Weyl scalars for all values of $p/M$ using Eq.~\eqref{eq: singauss}. Solid circles denote values obtained from the amplitude fit, while solid squares correspond to the sinusoid fit. Following the procedure shown in Fig.~\ref{fig: CHE-params}, the range of phase shifts has been extended to the range $(-2\pi,2\pi]$.}
    \label{fig: Merger-params}
\end{figure*}

In these plots, we can also see that the CHE emission is produced in a chirplike form, which is a desired result for this type of emissions~\cite{Morras:2021atg}. In the upper left plot, we see evidently that when the incidence angle is sufficiently small to create a DC where both emissions have short time intervals in between, both the merger and CHE portions of the Q-plot fuse together, creating a complex behavior that cannot be accurately described. This makes even more evident cause of the oscillations in the estimated parameters for smaller $\theta/\theta_0$ in Figs.~\ref{fig: Merger-params} and~\ref{fig: Ring1-params}, especially in the phase shift, as explained in the following section of the Appendix~\ref{app: pars_merger_ringdown}.

\begin{table*}[htbp]  
\caption{Calculated fitting values with respect to $\theta/\theta_0$ for the estimated parameters of the CHE portion of the Weyl scalar according to the methodology described in Sec.~\ref{Subsec: CHE}. In particular, the types of fits we use for each parameter are either ``Linear'' or ``Exponential'' [see Eqs.~\eqref{eq: linear trend} and~\eqref{eq: expo. decay}]. \c{The procedure to find the amplitude-related parameters and then the frequency and phase shift are thoroughly explained in Sec.~\ref{Subsec: Study of the Weyl}}. The uncertainties of the estimated coefficients are written on the last digits(s) at $1\sigma$ in parentheses.}
\centering 
\renewcommand{\arraystretch}{1.5}
\setlength{\tabcolsep}{3.3pt}
\begin{tabular}{|c|c|c|c|c|c|}
\hline
CHE parameter & Fit & $p/M$ & $\alpha$ & $\beta$ & $\gamma$ \\
\hline
\multirow{6}{*}{$A_{\rm CHE}$} & \multirow{6}{*}{Linear} & $0.095$ & $-0.6\c{3}(1)$ & $0.6\c{67}(9)$ & ... \\
\cline{3-6}
                        &                              & $0.1225$ & $-0.65(2)$ & $0.70(2)$ & ... \\
\cline{3-6}
                        &                              & $0.245$ & $0.3(2)$ & $0.1(2)$ & ... \\
\cline{3-6}
                        &                              & $0.3675$ & $0.8(2)$ & $-0.6(2)$ & ... \\
\cline{3-6}
                        &                              & $0.49$ & $1.5(2)$ & $-1.30(2)$ & ... \\
\cline{3-6}
                        &                              & $0.75$ & $1.0(1)$ & $-0.7(1)$ & ... \\
\hline
\multirow{6}{*}{$\sigma_{\rm CHE}$} & \multirow{6}{*}{Exponential} & $0.095$ & $0.0\c{5}(1)$ & $1\c{49(6)}$ & $10.8(\c{1})$ \\
\cline{3-6}
                        &                              & $0.1225$ & $0.08(1)$ & $192(7)$ & $9.99(7)$ \\
\cline{3-6}
                        &                              & $0.245$ & $0.08(2)$ & $370(23)$ & $8.73(7)$ \\
\cline{3-6}
                        &                              & $0.3675$ & $0.35(4)$ & $226(11)$ & $8.34(7)$ \\
\cline{3-6}
                        &                              & $0.49$ & $0.37(6)$ & $238(16)$ & $9.13(9)$ \\
\cline{3-6}
                        &                              & $0.75$ & $0.42(7)$ & $136(9)$ & $9.9(1)$ \\
\hline
\multirow{6}{*}{$t_0^{\rm CHE}$} & \multirow{6}{*}{Linear} & $0.095$ & $-\c{345(55)}$ & $\c{631(53)}$ & ... \\
\cline{3-6}
                        &                              & $0.1225$ & $-384(42)$ & $621(34)$ & ... \\
\cline{3-6}
                        &                              & $0.245$ & $-382(34)$ & $531(33)$ & ... \\
\cline{3-6}
                        &                              & $0.3675$ & $-302(19)$ & $426(19)$ & ... \\
\cline{3-6}
                        &                              & $0.49$ & $-313(17)$ & $429(18)$ & ... \\
\cline{3-6}
                        &                              & $0.75$ & $-231(12)$ & $344(13)$ & ... \\
\hline
\multirow{6}{*}{$\omega_{\rm CHE}$} & \multirow{6}{*}{Linear} & $0.095$ & $-1.\c{20(8)}$ & $1.\c{47(8)}$ & ... \\
\cline{3-6}
                        &                              & $0.1225$ & $-0.87(4)$ & $1.19(4)$ & ... \\
\cline{3-6}
                        &                              & $0.245$ & $0.05(26)$ & $0.4(2)$ & ... \\
\cline{3-6}
                        &                              & $0.3675$ & $-0.41(4)$ & $0.91(3)$ & ... \\
\cline{3-6}
                        &                              & $0.49$ & $-0.54(8)$ & $1.05(8)$ & ... \\
\cline{3-6}
                        &                              & $0.75$ & $-0.52(2)$ & $0.98(2)$ & ... \\
\hline
\multirow{6}{*}{$\phi_{\rm CHE}$} & \multirow{6}{*}{Exponential} & $0.095$ & $0.\c{09(2)}$ & $\c{112(8)}$ & $\c{-0.9}(1)$ \\
\cline{3-6}
                        &                              & $0.1225$ & $0.15(3)$ & $132(10)$ & $\c{-0.63}(9)$ \\
\cline{3-6}
                        &                              & $0.245$ & $0.20(4)$ & $215(18)$ & $\c{0.68}(7)$ \\
\cline{3-6}
                        &                              & $0.3675$ & $0.51(4)$ & $141(7)$ & $\c{1.57}(5)$ \\
\cline{3-6}
                        &                              & $0.49$ & $0.59(7)$ & $138(9)$ & $\c{2.67}(8)$ \\
\cline{3-6}
                        &                              & $0.75$ & $0.73(9)$ & $75(5)$ & $\c{2.7}(1)$ \\
\hline
\end{tabular}
\label{Tab: Annex}
\end{table*} 

\section{Figures of estimated parameters for the merger and ringdown}
\label{app: pars_merger_ringdown}

\begin{figure*}[htbp]
    \centering
    \includegraphics[width=\linewidth]{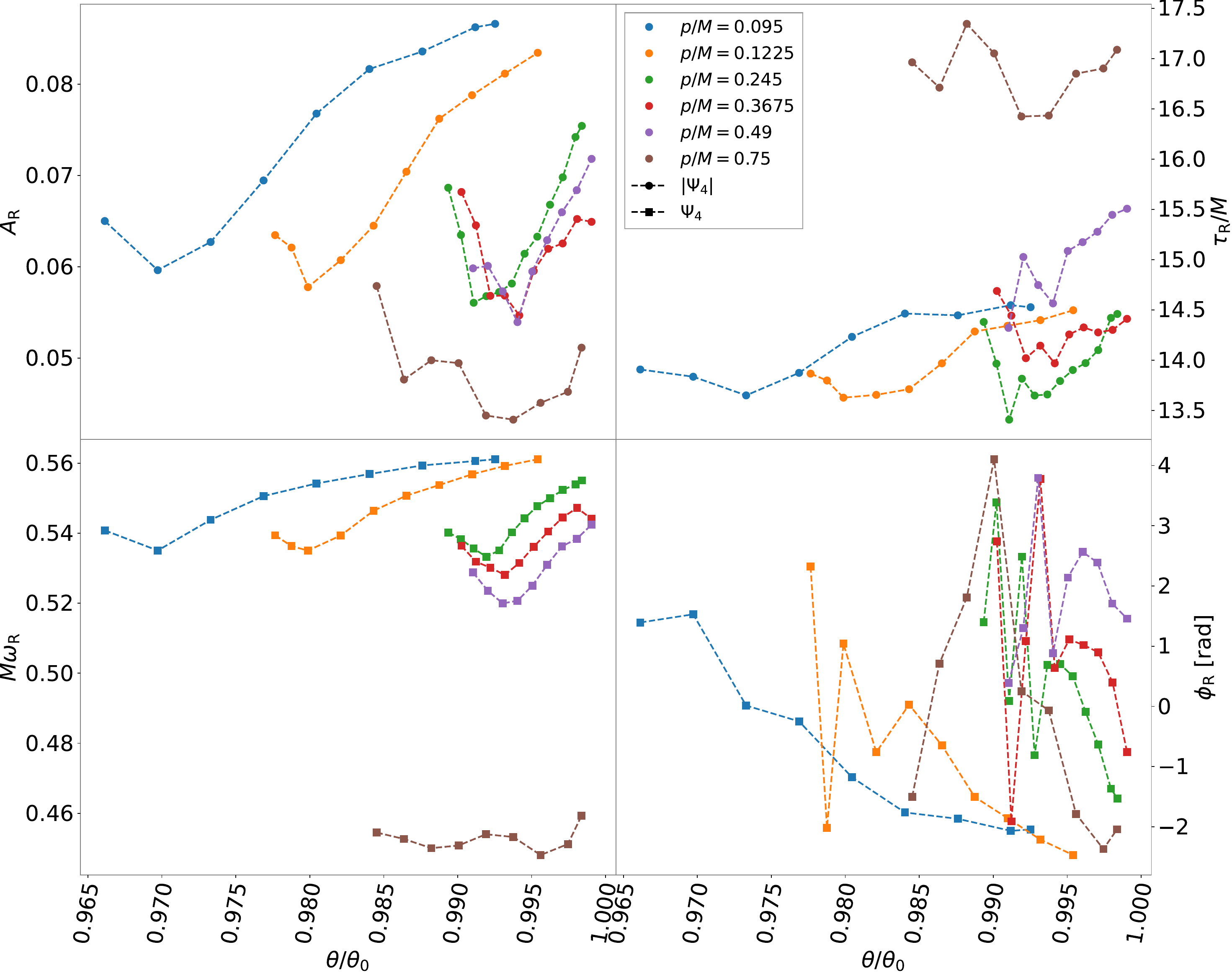}
    \caption{Estimated parameters of the first mode of the ringdown portion of the Weyl scalars for all values of $p/M$ using Eq.~\eqref{eq: ringdown}. Solid circles denote values obtained from the amplitude fit, while solid squares correspond to the sinusoid fit. Following the procedure shown in Fig.~\ref{fig: CHE-params}, the range of phase shifts has been extended to the range $(-2\pi,2\pi]$.}
    \label{fig: Ring1-params}
    \end{figure*}

In this Appendix, we show the obtained parameters for the merger emission of each case (Fig.~\ref{fig: Merger-params}). The procedure is explained in full detail in Sec.~\ref{Subsec: merger, ringdown}. We can see that general trends can be seen (especially for $t^M_0, \ A_{\rm M}$ and $\omega_{\rm M}$), but there is room for improvement. This is especially evident for the phase shifts, more specifically, for low $\theta/\theta_0$ in each scenario, where oscillations appear and no clear trends can be deduced. A very similar phenomenon occurs for the estimated ringdown parameters (Fig.~\ref{fig: Ring1-params}), where oscillations come in play once again.

As no clear trends have been found in this work, we have not carried out any fits to the parameters. We will attempt to obtain clearer trends in future works, as well as fitting the different parameters with respect to $\theta/\theta_0$, as it was done for the CHE case in Sec.~\ref{Subsec: CHE}, with Eqs.~\eqref{eq: linear trend} and.~\eqref{eq: expo. decay}. Nevertheless, this study helps us see the orders of magnitude that the GW emissions in DC events have, paving the way towards a future DC waveform model.

\begin{figure}[t!]
    \centering
    \includegraphics[width=\columnwidth]{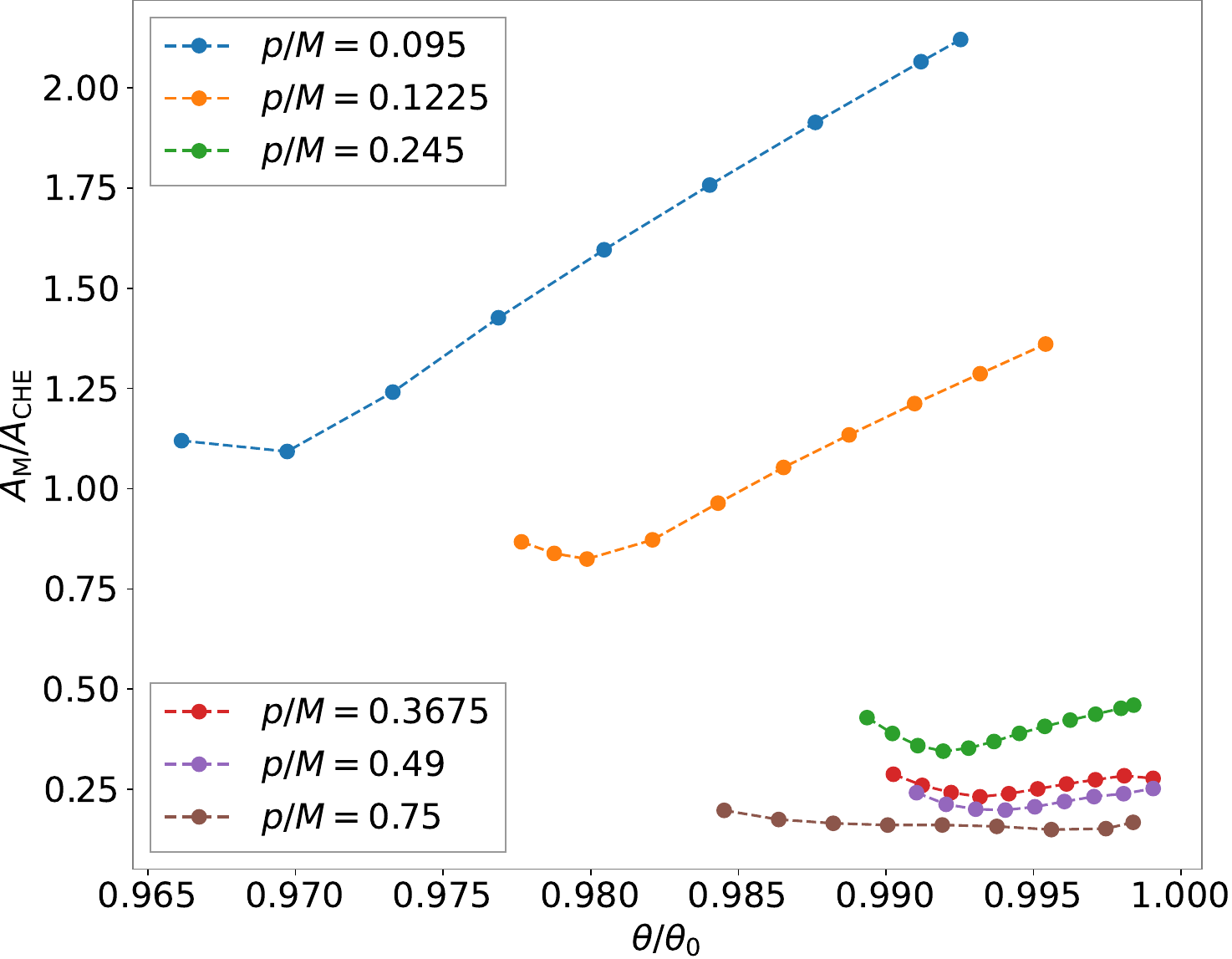}
    \caption{Ratio of the estimated amplitudes for the merger and CHE emissions ($A_{\rm M}/A_{\rm CHE}$), obtained using Eq.~\eqref{eq: singauss}. These values have been separately shown in Figs.~\ref{fig: Merger-params} and~\ref{fig: CHE-params}.}
    \label{fig: amp-prop}
\end{figure}

Furthermore, in Fig.~\ref{fig: amp-prop}, we plot the ratio $A_{\rm M}/A_{\rm CHE}$, which shows us that the least relativistic scenarios presented in this work have merger amplitudes ranging from about the order of the CHE emission to twice as large. Meanwhile, the most relativistic scenarios present much stronger CHEs, reaching values of CHE amplitudes four or five times larger than that of the merger. 

\section{An example of the overlap of the tails of both emissions}
\label{app: overlap}

\begin{figure}[t!]
    \centering
    \includegraphics[width=\columnwidth]{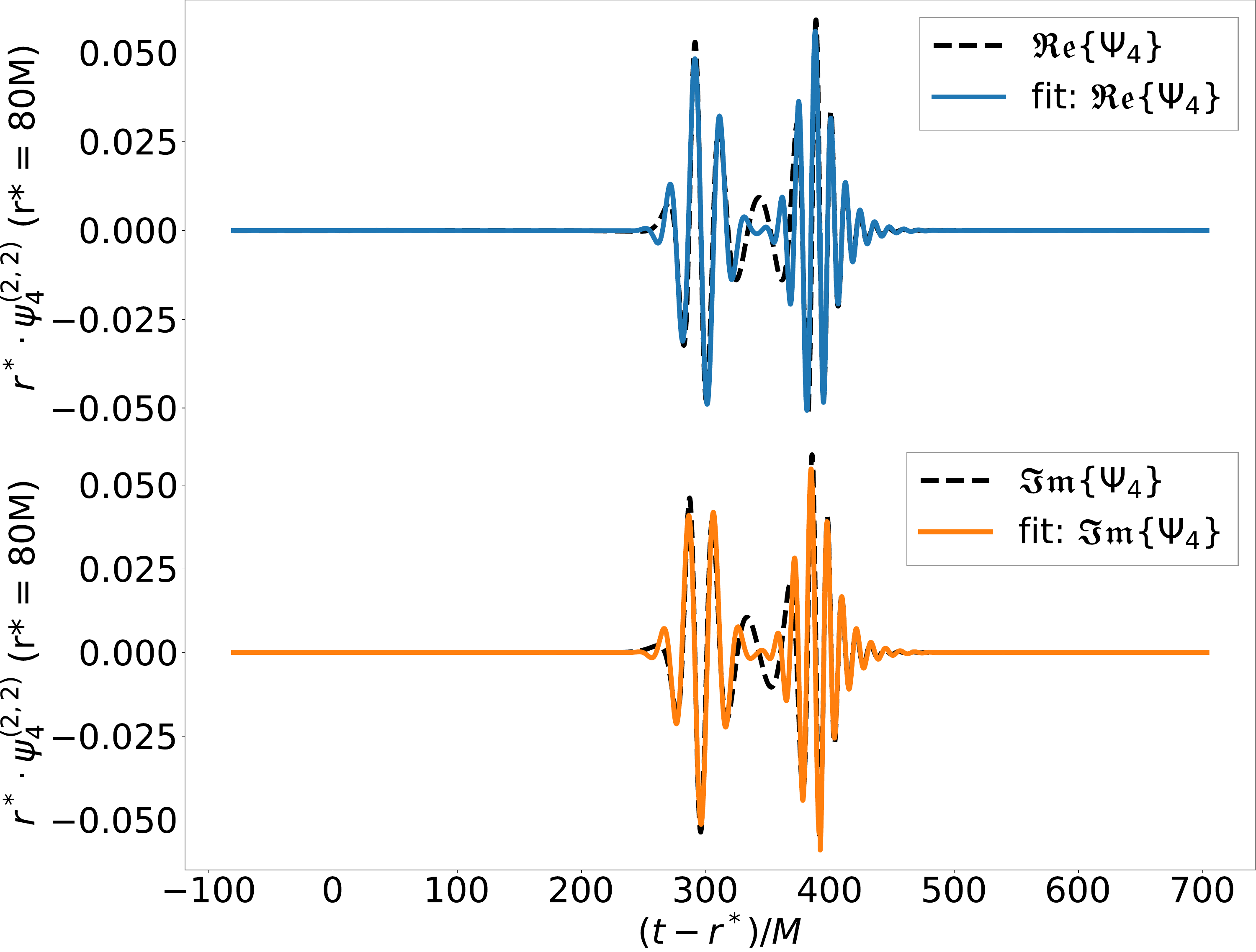}
    \caption{Real (blue) and imaginary (orange) \c{components} (upper and lower panels, respectively) \c{of the best-fit} along with the simulated signal (black, dashed) of the rescaled Weyl scalar at a detector at a distance $r^*=80M$ from the c.m. The initial conditions are $p/M=0.095$ and $\theta=6.2108^\circ$, with zero initial spins.}
    \label{fig: fit-psi-2}
\end{figure}

In Fig.~\ref{fig: fit-psi-2}, we show the performance of the combination of fits described in Sec.~\ref{Subsec: Study of the Weyl} for a case where the tails of both \c{gravitational-wave emissions} combine with each other. This is the main reason we observe the oscillations for the merger-ringdown estimated parameters (especially the phase shifts) in Figs.~\ref{fig: Merger-params} and~\ref{fig: Ring1-params}. 

These issues we observe at the tails for emissions with small time-interval might be potentially fixed by a time-dependent frequency or expansion of Eqs.~\eqref{eq: singauss} and~\eqref{eq: ringdown} at higher-order modes, and may be a possible source of future work for refining this phenomenological model.

\balance
\bibliography{main}

\end{document}